# Review: Representing and extracting knowledge from single cell data

**Authors:**

Ionut Sebastian Mihai[1,2,3], Sarang Chafle[1,2], Johan Henriksson[1,2*]

**Affiliations:**

1. Laboratory for Molecular Infection Medicine Sweden (MIMS), Umeå, Sweden

2. Umeå Centre for Microbial Research, Department of Molecular Biology, Umeå University, Umeå, Sweden

3. Industrial Doctoral School, Umeå University, Umeå, Sweden.

* Corresponding author

**Contact:**

Ionut Sebastian Mihai, ionut.sebastian.mihai@umu.se , https://orcid.org/0000-0002-9322-5879
Sarang Chafle, sarang.chafle@gmail.com, https://orcid.org/0000-0003-2475-3528
Johan Henriksson, johan.henriksson@umu.se, https://orcid.org/0000-0002-7745-2844

# Abstract

Single-cell analysis is currently one of the most high-resolution techniques to study biology. The large complex datasets that have been generated have spurred numerous developments in computational biology, in particular the use of advanced statistics and machine learning. This review attempts to explain the deeper theoretical concepts that underpin current state-of-the-art analysis methods. Single-cell analysis is covered from cell, through instruments, to current and upcoming models. A minimum of mathematics and statistics has been used, but the reader is assumed to either have basic knowledge of single-cell analysis workflows, or have a solid knowledge of statistics. The aim of this review is to spread concepts which are not yet in common use, especially from topology and generative processes, and how new statistical models can be developed to capture more of biology. This opens epistemological questions regarding our ontology and models, and some pointers will be given to how natural language processing (NLP) may help overcome our cognitive limitations for understanding single-cell data.

# Introduction

Single-cell datasets are among the most complex data currently generated, and the field is a major driver for new bioinformatic methods. Datasets can now encompass up to one million observations, and 20-50k measurements per cell, making it hard to even visualize the data. Furthermore, the data is incredibly noisy, and relies on our ability to detect individual molecules. To overcome the noise and extract a meaningful interpretation, it is usually not sufficient to look at individual cells. Instead, the data is fitted to increasingly advanced statistical models.

Statistics are commonly concerned with *data, or observations* (what we measure), a *model* (what generates the data), and underlying variables (frequently hidden and abstract in nature). In this review, we will be concerned with the nature of the data and how different models can help us explain it. In modern statistical language, the relationship between data and variables can be recast in the following general form:

$$\textbf{Single-cell observation}_i \sim \textbf{Model[latent variables}_i\textbf{]}$$

Here the model is the choice of the statistical distribution, parameterized by hidden (latent) variables. These abstract variables need to be given a meaning by the analyst, and a philosophical discussion cannot be avoided (discussed in the second part of this review). Most latent variables are for each observation (cell), while some variables may be in common for all observations or simply considered part of the model itself. The latent variable space typically has a lower dimension than the data space (Figure 1a), representing the aggregation of knowledge. Both Frequentist and Bayesian statistics are in use, with Bayesian models becoming increasingly popular. Their strength lies especially in their ability to model complex noise, which arises from both the biology and the technical measurement. Bayesian models furthermore support updating (i.e. adding data to a previously fitted model), avoiding complete recomputation as new evidence surfaces.

There is usually not an obviously “correct” choice of model for single-cell data, but model choice can be motivated by a hypothesis of the nature of the data. However, one hypothesis can correspond to multiple models, and multiple models can correspond to one hypothesis (Gelman and Hill 2006) (Figure 1b). For example, the abundance of RNA in a tube can increase both due to the number of molecules increasing, as well as if the gene switches to a longer isoform (Figure 1c). How this affects RNA-seq depends on the precise chemical details in the library preparation, which is why those are covered extensively in this review. The conceptual difference between hypotheses and models is also especially important to be aware of in the context of hypothesis testing, where the fit of one model is compared to another (Figure 1d). Another way of motivating models is in terms of their interpretability - a model might fit the data very well, but because of its complexity, it might be hard to interpret. Finally, model choice can be motivated in terms of how well a model fits to not-yet-seen data. The machine learning (ML) field has especially emphasized this aspect and provided tools such as cross-validation and penalization to improve upon it. It is now even possible to fit models for which classically there would not be enough data given the number of parameters. These general topics of statistics and ML are beyond the scope of this review.

This review is organized around our cognitive view of modeling (Figure 1e). Beyond the general form equation, it will also cover how single-cell data is physically generated, how data is preprocessed, and how the latent space can be interpreted from topological and biological standpoints.

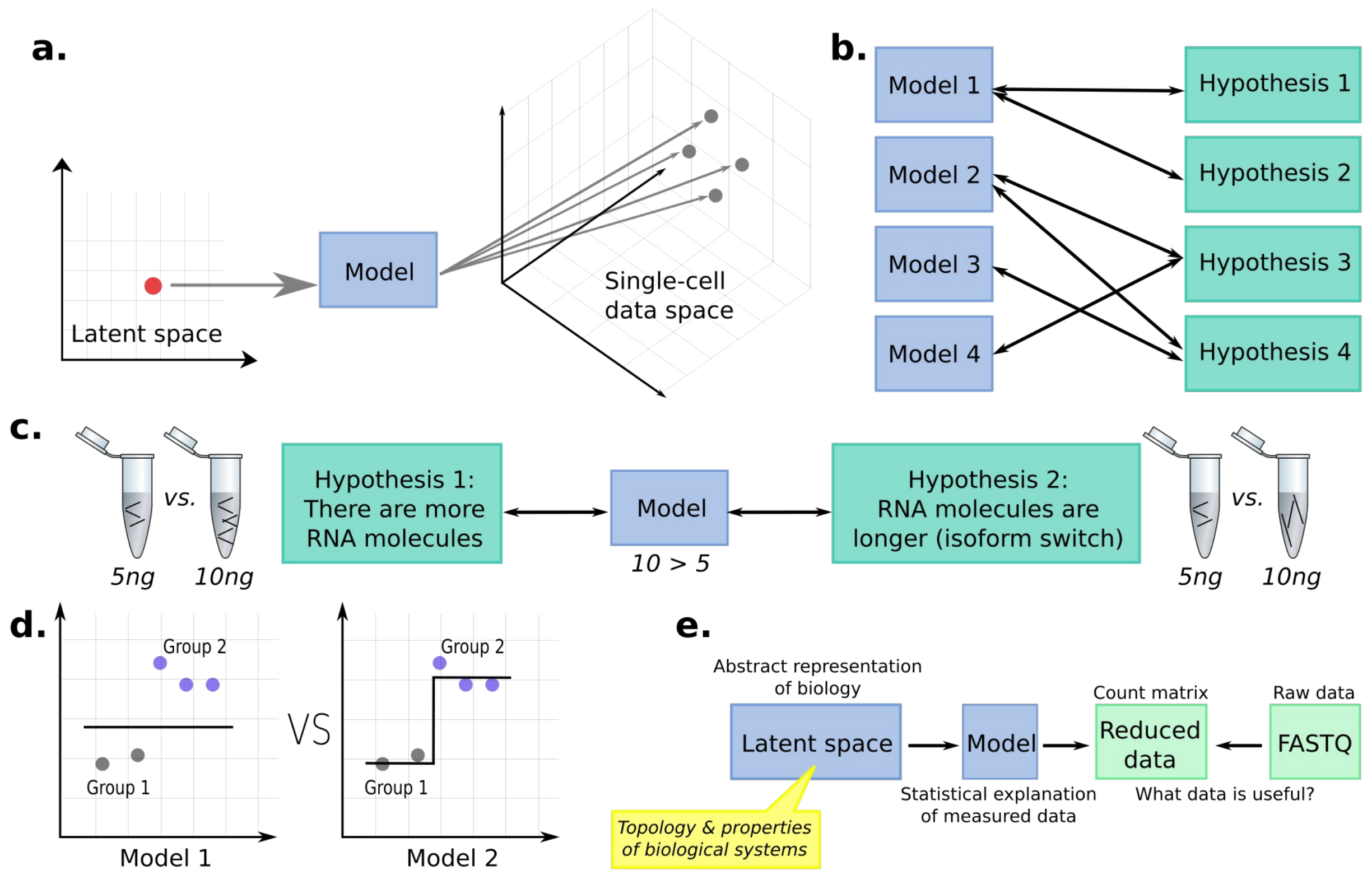

**Figure 1: Introduction to single cell data representation. (a)** A smaller space of latent variables can through a model represent a larger space of single-cell observation data. Several data points can be represented by the same point in the latent space. **(b)** One hypothesis can correspond to multiple models, and multiple models can correspond to one hypothesis. **(c)** Two hypotheses about the abundance of RNA in a tube lead to the same statistical model. **(d)** Hypothesis testing can be done by comparing the fit of one model to another; here, the left model corresponds to no difference between the samples, while the right model enables there to be a latent difference between group 1 and 2. **(e)** The data modeling procedure from the perspective of this review.

## The standard analysis pipeline

In the beginning of single-cell analysis, tools were reused from bulk RNA-seq, such as the Bowtie2 aligner (Langmead and Salzberg 2012), and DESeq2 (Love et al. 2014) for differential expression. Single-cell is however fundamentally different in that there are many more samples (cells) compared to bulk (tissue averages). Thus techniques from unsupervised ML were borrowed to aid visualization and comparison. While this worked well, it required a rather skilled bioinformatician. This has led to the development of several user-friendly R and Python packages, which streamline the historically most common operations. Among the most well-known such packages are Seurat (Satija et al. 2015), Signac (Stuart et al. 2021), monocle (Cao et al. 2019) and ArchR (Granja et al. 2021) for R, and Scanpy (Wolf et al. 2018) for Python.

The standard pipeline (Figure 2a) proceeds as follows: **(1)** Alignment of sequencing data to a reference genome, **(2)** Gathering data for each cell, **(3)** Reducing the sequencing data into per-cell features, e.g., gene expression levels, transcription factor binding level or enhancer accessibility, **(4)** Further quality control with doublet removal and feature selection, **(5)** Dimensional reduction and clustering, **(6)** Comparison of cells and clusters. In every step, data is removed (Figure 2b). Ideally, there would only be one step, but it would be too slow to be practical. Thus, in fact, there are several data representations, and the process should be seen as a funnel to a *p*-value or plot. Each data representation thus needs to contain enough information to be reduced into the next representation (e.g. the normal count reduction makes it impossible to distinguish the scenario outlined in Figure 1c in later steps). This review thus inevitably covers most steps in the analysis process. One aspect not covered is that there are specific file formats for the intermediate representations (e.g. Anndata (Wolf et al. 2018), loom, https://github.com/mojaveazure/loomR, or Arrow (Granja et al. 2021)), which limits one from easily making changes in these representations.

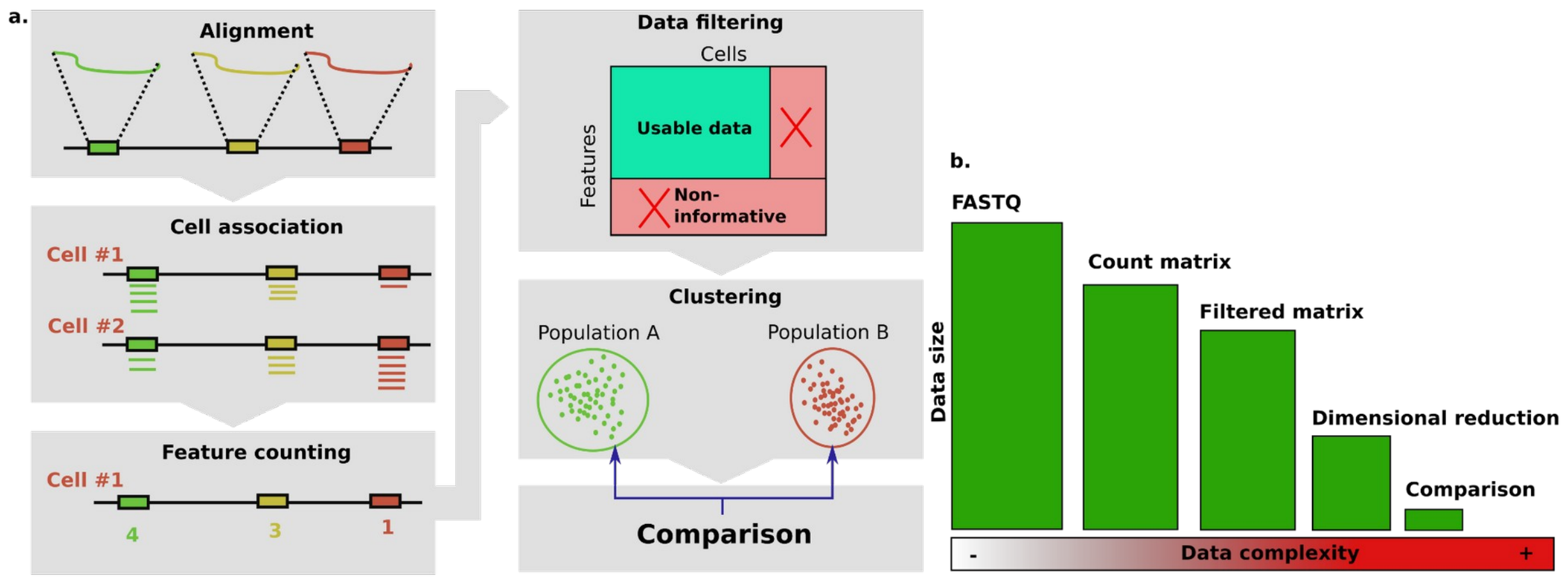

**Figure 2: The standard analysis pipeline. (a)** All the common steps in the analysis of RNA-seq or ATAC-seq data. **(b)** Information is reduced, and lost, in every step of the analysis; but the value or complexity of the remaining information increases.

# The archetype single-cell chemistries

Before delving into how data is represented, one must understand how the data arises and to what it physically corresponds. Two archetype library preparation methods (RNA-seq and ATAC-seq), colloquially called "chemistries", will be covered in this section. These chemistries can now be performed on the same input cell, enabling close comparison of these modalities and sharing of the latent space (Argelaguet et al. 2018; Gayoso et al. 2021). The principles behind the multiome protocols (Lee et al. 2020) (simultaneously measuring more than one biological aspect of the same cell) are virtually the same as for RNA-seq and ATAC-seq separately.

## Single-cell RNA-seq

scRNA-sequencing is a method to quantify which genes are transcribed. The focus is normally on mature mRNA, which at the end contains a 5' nucleotide cap and a 3' poly adenine (3' Poly-

A) tail. Furthermore, introns will at some point be spliced out. Because the mRNA levels enable such broad interpretation of the function and behavior of a cell at a given moment, RNA-seq has become the workhorse for much biological exploration. scRNA-sequencing is also the basis of more complex protocols, such as single-cell ribo-seq (VanInsberghe et al. 2021), which can tell where ribosomes are located. RNA-seq was the first omics protocol miniaturized for single-cell applications, with multiwell plate Smart-seq2 (Picelli et al. 2013) becoming the most popular protocol. Other competing protocols existed; for example CEL-seq2 (Hashimshony et al. 2016) only captured the 3' part, while STRT-seq (Islam et al. 2012) captured the 5' part (in the idealized scenario). The first step in any protocol is reverse transcription (RT). Because of the abundant ribosomal RNA, virtually all protocols so far have used oligo-dT RT-primers which bind to the 3' polyA of the mRNA. However, oligo-dT primers can also bind anywhere inside the RNA but with lower efficiency, typically in stretches of 3xA or more (Kozak 1991) (which is essential for estimating "RNA velocity" (Bergen et al. 2021) from the quantification of unspliced mRNA that would not be included if only 3' polyA mRNA was included).

Almost all single-cell protocols require early addition of flanking PCR handles, which enable a first PCR (pre-amplification) using cDNA-sequence-independent PCR primers. The first PCR handle is included in the RT-primer. A key innovation is the use of template switching, where RT can continue from the 5' end of the RNA to another oligo. This preferentially occurs if the RNA has a 5' cap, and if the RT enzyme adds additional nucleotides past the RNA, which can be made complementary to an oligo ("the template switching oligo, TSO"). The TSO can be designed to carry a second PCR handle. With known common 5' and 3' flanking sequences, the first PCR is then easily designed (Figure 3a). To aid in read deduplication (discussed later), this is also the step in which a Unique Molecular Identifier (UMI) is introduced, as a stretch of random nucleotides in either the RT primer or TSO.

Next, all libraries aimed for Illumina short-read sequencing require fragmentation of the cDNA down to sizes that can bind to the flow cell (below 1.5kb, ideally 700bp). Smart-seq uses Tn5, an enzyme that simultaneously fragments the cDNA and adds new PCR handles (Figure 3b). This protocol is simple, but the ends of input DNA are lost unless 3'/5' adapters are added somehow (such as by extended TSO and RT primers in Smart-seq3 (Hagemann-Jensen et al. 2020)). Furthermore, because Tn5 adds s5/s7 tags randomly, 50% of the fragments being s5-s5 or s7-s7 will be suppressed in the PCR after tagmentation. As an alternative to Tn5, some protocols instead use enzymatic shearing, dA-tailing and sticky adapter ligation (Figure 3c). CEL-seq2 (Hashimshony et al. 2016) and STRT-seq (Islam et al. 2012) also fragment the cDNA, but targets either the 5' or 3' PCR handle from the cDNA preparation to enrich for the corresponding fragments. This is possible only if different adapter sequences are used for 5' and 3' respectively (Figure 3d). After a second PCR, the fragments will contain cDNA, library indices, and sequences that will bind to the Illumina flow cell (Figure 3e).

Single-cell protocols can also be performed with microfluidics, where a cell is encapsulated in a droplet containing enzymes, buffer and a bead with oligos (Klein et al. 2015; Macosko et al. 2015). These protocols have severe constraints: buffers cannot be changed, adding liquid to droplets is challenging, and somehow each cell has to obtain a unique library index. For multiwell plates, different index oligos are added to each well, and the well-index relation is known. Droplets instead depend on beads having oligos that carry a random library index (in this context called the “cell barcode”), with sufficiently many indices such that no cells obtain the same index (Figure 3f). Because it is hard to add liquid to droplets, PCR is rather performed after droplets have been de-emulsified and pooled. This limits the possibility of adding the cell barcode to the RT-step. Because the barcode can only be added to the 3’ or 5’ of the cDNA (RT-primer or TSO), only these parts of the cDNA are normally sequenced in the 10x system (Figure 3g).

Finally, split-and-pool-type protocols should be briefly mentioned (Vitak et al. 2017; Rosenberg et al. 2018). These use the cell itself as the “droplet”, by careful permeabilization. The cell barcode is made up of a combination of oligos, with each extra oligo added after pooling the cells and splitting them into a new 96/384 well plate. These protocols are less constrained than microfluidic droplet based protocols, but overall share the limitation of only capturing 3’ or 5’ RNA. Split-and-pool protocols can also be implemented using the 10x chromium (Datlinger et al. 2021).

While this section only covers a fraction of the chemistries ever made, knowing this much about RNA-seq is sufficient to be able to statistically model most of existing data. Recent advances in chemistry will however introduce new statistical challenges, e.g, there are alternatives to oligo-dT RT. For example, microSPLiT uses *in vitro* polyadenylation to also capture bacterial mRNA efficiently (Kuchina et al. 2021). Similarly, this can be done on eukaryotic fragmented mRNA, enabling full-RNA capture also in the 10x droplet system (as in VASA-seq (Salmen et al. 2022)). Because Tn5 also digests RNA:DNA hybrids, the PCR before tagmentation can also be avoided (Di et al. 2022; Xu et al. 2022). As a benefit, the fragmentation sites for each input RNA molecule is virtually unique, simplifying deduplication and the relation between reads and input RNA molecules.

## Single-cell ATAC-seq

ATAC-seq (Assay for Transposase-Accessible Chromatin using sequencing) is a method that assesses which regions of the chromatin are accessible (“open”) using the transposase Tn5 (Yan et al. 2020). This technology relies on the assumption that inactive transcribing (and replicating) regions of the DNA tend to be highly compacted around histones (“closed”), while active regions are not. When Tn5 tagmentation is done on genomic DNA (gDNA), before purification, the fragmentation patterns will thus be dictated by gDNA-binding proteins, marking unshielded gDNA as open (Figure 3h). Since tagmentation is well-suited for small-input single-cell applications (Buenrostro et al. 2015), it has been adapted to also enable scChIP-seq (e.g. single-cell CUT&Tag (Bartosovic et al. 2021)). By adding a H3K9me3-targeting chromodomain to the Tn5 enzyme, both open and closed regions can be assessed in parallel, enabling the measurement of “chromatin velocity” (Tedesco et al. 2022). scATAC-seq is thus the basis for a family of related protocols.

Because mitochondrial DNA (mtDNA) is abundant and highly accessible, a key first step in ATAC-seq is the extraction of the nuclei. This step can however be relaxed for niche protocols that use somatic mutations in the mtDNA for lineage tracing purposes (Ludwig et al. 2019). During nuclei extraction, the nuclei are also permeabilized, giving access to the gDNA. Because Tn5 binding is separate from the subsequent gDNA fragmentation, which is induced by heating or denaturing agents, Tn5 can be added to nuclei in bulk prior to nuclei single-cell separation (Chen et al. 2018). The individual separation of nuclei can then be done both in multiwell plates as well as in microfluidic droplets. A PCR is finally performed to attach library indices, or cell barcodes in the case of droplets. This is thus unlike 10x chromium droplet RNA-seq, where cell barcodes are instead added by RT.

## Comparison of readout methods

One major difference between RNA-seq and ATAC-seq is the number of possible molecules overlapping the same genomic region (Figure 3i). In ATAC-seq, assuming a diploid genome, there can be at most 2 fragments. For RNA-seq, there is however no limit on the amount of RNA for one gene. Furthermore, more than one RT primer can bind to one RNA molecule, even if oligo-dT primers are used. Thus the UMI represents the number of RT events, not the number of RNA molecules.

Another difference is that there are more steps in the RNA-seq chemistry. Assuming that one RNA molecule results in one cDNA molecule, the subsequent fragmentation step still produces more than one sequenceable fragment per RNA molecule. This is not just the case for Smart-seq2, but also for 5’ and 3’ capture methods, which produce overlapping fragments of different lengths (Figure 3j). This must be kept in mind during UMI-based deduplication. In principle, the overlapping fragments can be combined into one single read based on the UMI, but the authors do not know of any such software.

Based on RNA-seq and ATAC-seq, additional steps can lead to new types of readouts. Metabolic RNA labeling protocols to measure rate of transcription (Qiu et al. 2020) are effectively RNA-seq protocols. Protein levels can be measured by using oligo-tagged antibodies, thus turning protein detection into a sequencing problem. CITE-seq and REAP-seq are two ways in which antibody-attached oligos enable detection by RNA-seq protocols (Stoeckius et al. 2017; Peterson et al. 2017; Mimitou et al. 2019) (Figure 3k). As an alternative, tagged lipids (such as 10x genomics CellPlex) can label a variety of cells. All of these follow the chemistry expected for RNA-seq, except since the fragments are already sufficiently small to fit Illumina sequencers, no fragmentation is needed. Instead, a separate enrichment PCR is sufficient to selectively extract them from the pre-amplified cDNA library. Labeling of cells can also be done in an ATAC-seq-compatible manner through a different oligo design (Mimitou et al. 2021). In addition, the ATAC-seq protocol can be modified for detection of CRISPR sgRNAs (Pierce et al. 2021), and for quantification of other genomic regions (unpublished).

To the author's knowledge, there are only two fundamentally different archetype readouts not covered by this review. Single-cell HiC generates a map of location-to-location abundances (Nagano et al. 2013; Stevens et al. 2017; Zhang et al. 2022). The other readout is single-cell whole-genome sequencing (Gawad et al. 2016). These are both rather niche protocols, requiring complex statistics tailored for the purpose, and thus not covered in this review.

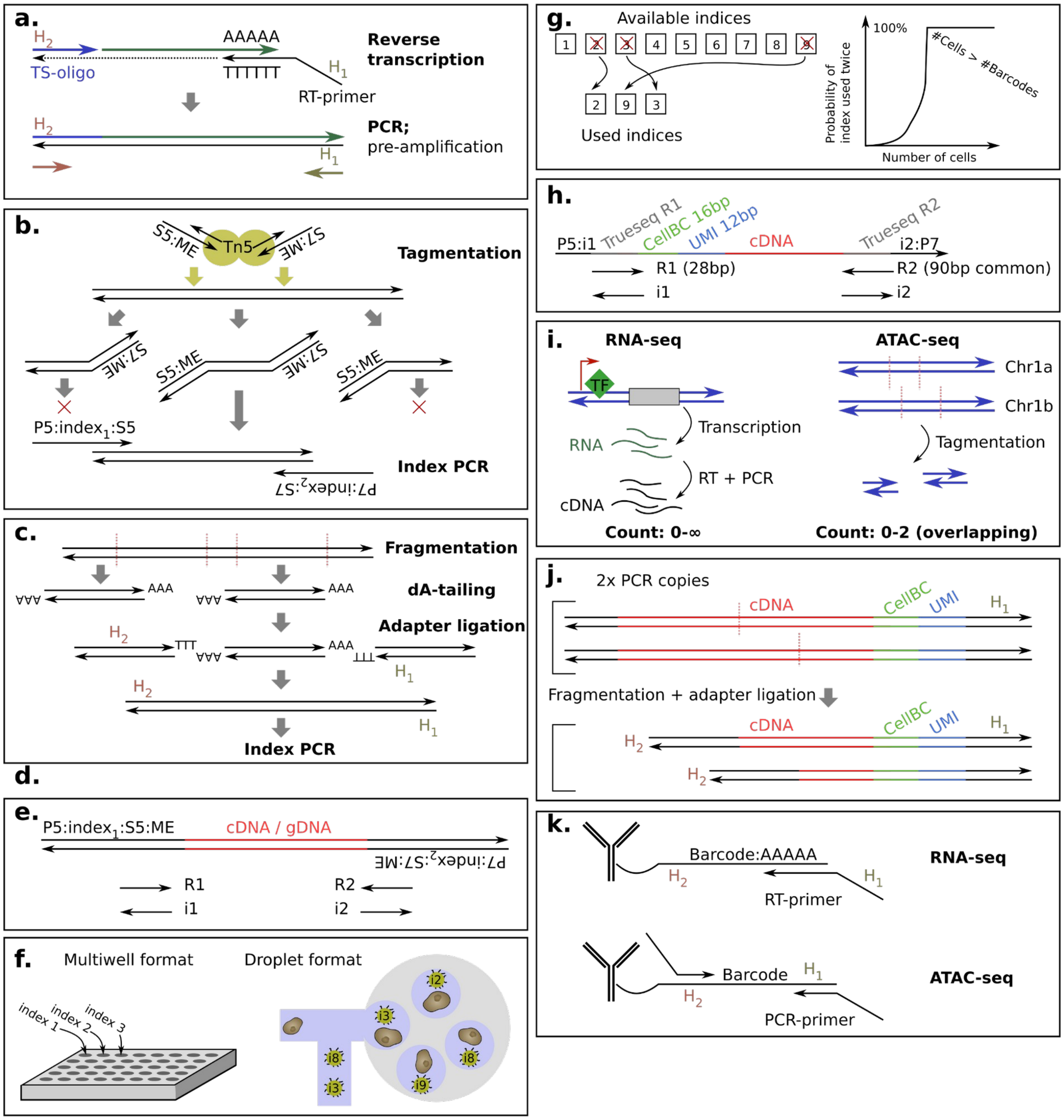

**Figure 3. The archetype single-cell chemistries. (a)** During reverse transcription, any choice of flanking sequences $H_1$ and $H_2$ can be added to 5' with template switching, and 3' by an extended RT primer. This aids subsequent PCR. **(b)** Tagmentation, fragmentation and addition of flanking tags by Tn5. The 5'- and 3' flanking parts will only be tagged on one side and lost in subsequent PCR. **(c)** Addition of adapters by enzymatic digest, dA-tailing and ligation. **(d)** Enrichment for 5' or 3' fragments can be done by PCR if the PCR for H2-H2 is suppressive (e.g. Nextera S5-S5 or S7-S7). **(e)** A typical final library DNA molecule ready for sequencing by Illumina short read sequencers. **(f)** Multiwell protocols have per-well defined cell barcodes, while droplet protocols rely on random cell barcodes. **(g)** The chance of picking the same random index twice depends on the number of available indices as compared to how many indices are used. **(h)** The final 10x chromium RNA-seq library structure differs from a typical Illumina library. **(i)** The expected number of overlapping fragments differ between RNA-seq and ATAC-seq. **(j)** cDNA fragmentation can cause overlapping fragments because of the pre-amplification. **(k)** Proteins can be measured by RNA-seq and ATAC-seq protocols using oligo-tagged antibodies.

# From single-cell chemistry to statistics

## Sequencing data preprocessing and initial data reductions

Independently of the chemistry used, the first steps of alignment and barcode-to-cell association are largely the same. If the data stems from a multiwell experiment, then classical bulk RNA-seq or ATAC-seq tools have commonly been used. For droplet data or more complex chemistries, dedicated tools exist that scale better for large datasets. CellRanger operates on 10x chromium data, but can be replaced by the faster and more flexible Star SOLO (Blibaum et al. 2019) pipeline (which is also suitable for multiwell plates). The output alignment will, for each sequencing read, contain information about position (chromosome name, from, to), and sequence differences *vs* the reference genome (Figure 4a). This is more information than is commonly used, and to speed up later algorithms, most information is filtered/reduced in a manner that depends on the needs of subsequent steps.

For RNA-seq, only information about which gene the read overlaps is typically retained. Any UMI is used to further deduplicate reads. The result is counts of fragments per gene and per cell, called the count table. Some computational methods retain more information about the read. RNA velocity, for example, stores if the read is intronic or exonic, and thus reduces the reads to two counts per gene and cell. Interestingly, the 10x software Cellranger only counts exonic reads by default, but can optionally count everything (recommended for single-nuclei protocols). More information exists, including about isoform usage, but novel computational approaches need to be developed to make better use of this.

ATAC-seq analysis is more challenging because, unlike for genes, there is no accepted “list of enhancers”. Instead, this list is defined for each dataset by collecting fragments across all cells and performing peak calling in a manner identical to bulk ATAC-seq or ChIPseq. MACS2 (Zhang et al. 2008) is commonly used, while Cellranger has its own algorithm. With peaks defined, it is then possible to collect fragment counts per peak. It is an open question as to what other information can be extracted; after later analysis steps, it is possible to reanalyze the raw reads to detect transcription factor (TF) binding sites (“TF footprinting” (Bentsen et al. 2020)). This shows that in some cases, even if information is not retained in a reduction, it may be possible to backtrack to the raw data to extract further information.

It is sometimes possible to skip the slow alignment step and immediately count the feature overlaps. Alevin (Srivastava et al. 2019) does so for RNA-seq by instead solving a k-MER deconvolution problem over all reads. This requires a reference of expected sequences of high confidence, which cannot be fulfilled for variable regions such as the T cell receptor (TCR). This approach is however particularly promising, as it avoids the problems of non-uniquely aligned reads, and the counting speed is sufficiently high that downstream statistics can be done by bootstrapping. One can thus expect a future single-cell pipeline that is entirely based on a k-MER sequencing data representation instead of the current count matrices.

## Filtering cells and features

The resulting count table is, in more general terms, said to consist of quantified “features” for each cell. To speed up computation and avoid false positives, the number of features can be reduced (Figure 2a panel 4). For example, if a gene is expressed equally in all cells, then it does not contribute much information. In the early days, ERCC (External RNA Controls Consortium) spike-in RNA was used to estimate technical variance. However, these days, features are selected by pure comparison of their dispersion to other similar features. For gene expression, this means other transcripts of similar abundance (Brennecke et al. 2013). Genes having less variance than the expected technical variance can also be ignored.

Another useful reduction is to remove observations of low quality. Cells for which there are few reads (low coverage) do not contribute much information and are usually removed based on a lower cut-off (no gold standard exists). There may also be free floating RNA or DNA, which may enter droplets of other cells (background). The background can be modeled and its effect removed to a certain extent (Young and Behjati 2020), but cells with few reads are particularly vulnerable to this bias. RNA also sticks to cells, and this may cause bias if it preferentially sticks to neighboring cells, since cells are not randomly distributed in the tissue. Finally, more than one cell can enter a library, especially if droplet microfluidics is used. Such droplets can be detected by ML after simulating the mixing of all cells. Several such packages exist (Wolock et al. 2019).

## Unique molecular identifiers (UMIs) and barcode correction

Because of the limited amount of input DNA/RNA, after PCR, it is highly likely that some fragments will be sequenced more than once. To avoid double counting, some fragments are equipped with unique molecular identifiers (UMIs). These are simply stretches of N-nucleotides (random mix of ATCG), with a suitable length depending on the expected number of duplicate fragments. If two fragments are equal, and also share a UMI, then they are assumed to have the same origin (Figure 4b). Removal of extra copies is called deduplication. UMIs cannot be attached during PCR, but rather only at steps which can only happen once: ligation, reverse transcription and template switching. In Smartseq2-style protocols, it is impossible to add UMIs representing RT-events for all final fragments, as fragmentation by Tn5 separates most inner cDNA fragments from 5’ or 3’ UMIs. Smartseq3 has, however, also added UMIs to the 3’ and 5’-most fragments (Hagemann-Jensen et al. 2020).

The handling of UMIs is generally a matter of preprocessing, as is done by Cellranger. Because sequencing errors can occur, UMIs may need to be bioinformatically corrected. If a small number of reads contain UMIs similar to highly abundant UMIs, then reads can be assumed to be due to sequencing errors (Smith et al. 2017). If many of the possible UMIs are used, then the assumption of UMIs being unique breaks down, and it may be necessary to treat them with more advanced statistics. This is because the birthday paradox (Wikipedia contributors 2022) - the chance of at least two UMIs being shared, if picked randomly from a large pool - increases surprisingly fast. The details are not covered here, but it is important to be aware of the problem such as to design UMIs and experiments accordingly.

Cell barcodes in droplet data can be corrected similarly to UMIs, but with higher confidence if the random sequence comes from a predefined list (called a whitelist). In this case, there are several algorithms enabling the design of oligo sequences that can be corrected efficiently (Buschmann and Bystrykh 2013). As an example, the 10x chromium RNA-seq chemistries use whitelists of up to 1.4M cell barcodes.

## Toward a statistical model

In early days of model fitting, little or no regard was given to statistical distributions. For simplicity, the euclidean distance (data *vs* fitted value) was commonly minimized, which in hindsight was a good choice - it actually has strong links to the normal distribution. The normal distribution is frequently a correct choice because it arises naturally for any variable that is the average of several other stochastic processes - a result denoted the central limit theorem. In biology, such averaging is common (e.g. a phenotype is usually the total result of many interacting genes). That said, the average might rather be on a log-scale. This happens if the results are multiplicative (e.g. one mutation increases length by 10%, and another mutation 10% on top, resulting in 1.1*1.1=1.21, which is more than 1+0.1+0.1=1.20 in the additive case). This is also implicit to any use of “fold change” in gene expression analysis, as it ignores the absolute gene expression level. Normal and log normal distributions are thus reasonably good and common choices to model biological processes. However, these are continuous distributions while sequencing data is discrete and has a rather different shape near zero (Figure 4c). Luckily, modern computing has enabled the use of more appropriate distributions.

The most important distribution for discrete sequencing data is the Poisson[λ] distribution. The Poisson distribution can be physically motivated to model the number of decays (happening with rate λ) under a certain time from a radioactive source (Figure 4d). This is because of an intrinsic physical property, namely that any radiative event is uncorrelated to when the last radiative event happened. In other words, it is a *memory-less process*. Such processes are widely modeled, even if not completely memory-less, because they are easy to handle mathematically. Sequencing can be thought of as a process of picking random DNA molecules from a semi-infinite tube. Even if the DNA has been PCR amplified, the probability of picking a copy of a previous molecule is approximately non-existent. This makes it approximately a memory-less process, where the Poisson rate parameter is dictated by the total number of reads, and how many percent of the molecules are expected to come from the gene/enhancer of interest.

The Poisson distribution is relevant to any sequencing context, but does not take biological properties into account. As such, it usually underestimates the variance. The Negative Binomial distribution, NB[rate, dispersion], is a natural extension that in addition to rate also has a dispersion (variance) parameter. It is a well-studied distribution that is the *de facto* standard for bulk RNA-seq analysis (Love et al. 2014). NB is also equivalent to a Poisson distribution, when the NB rate parameter in turn is Gamma distributed. Thus NB-distributed counts can be expected from a sequencer, if the continuous Gamma distribution represents the biological variation, and Poisson the sampling by sequencing. This makes the NB-distribution a first choice when analyzing data from new sequencing protocols. Various types of data normalizations can however also be applied, but are discussed and benchmarked elsewhere (Ahlmann-Eltze and Huber 2023).

## Statistics for RNA-seq

Single-cell data is highly noisy and begs for more complex models than bulk equivalents. The memory-less assumption behind Poisson is less appropriate because the pool of DNA is no longer semi-infinite. Rather, in our experience, up to 30% or more of the final DNA molecules from a 10x scRNA-seq experiment can be duplicated (calling for UMI-based deduplication). To motivate the best statistical models, it is thus necessary to understand details of the central dogma, how the final DNA molecules arise in the library preparation, and what physical properties thus can be expected.

It was noticed early on that RNA-seq counts for a gene follow a zero-inflated distribution (large number of 0-values, Figure 4e), starting a heated debate on the nature of single-cell data and whether there are biological reasons why some genes “drop out”. One study suggests that the zero-inflation problem is rather overrated, and that no zero-inflation is observed for ERCC spike-in RNA (Svensson 2020). However, it is easy to see that zero-inflation is a concern for plate-based full-length RNA-seq (Figure 4f). When a single RNA molecule is present, it can give rise to multiple counts; however, when the RNA molecule is not present, the count will be exactly zero. Our group has noticed that a single RNA molecule can give rise to multiple cDNA molecules, even in regular 10x droplet chemistry (unpublished). Statistically, the problem can also be seen as zero-inflation (dropout rate q) of count *C* (Poisson distributed):

$Count\ \ Z*C$, with $C\ \ Poisson[\lambda], p(Z=0)=q, p(Z=1)=1-q$

Or equivalently: $Count\ \ ZeroInflatedPoisson[\lambda, q]$

Letting the variable *Z* denote the presence of the RNA molecule, and *C* the cDNA count distribution during presence. However, this only applies to the case of a single RNA molecule, and as such, is likely to fit best to lowly expressed genes. Alternatively (and more correctly), the count can be seen as a sum of highly correlated variables. However, this treatment is difficult mathematically. A practical serious concern is identifiability (can the parameters be fitted given the data?): the ZINB model has three parameters (mean, dispersion, dropout rate), while NB has two (mean, dispersion) and Poisson only one (mean). If the dropout rate cannot be reliably fitted, then this might affect later steps, such as differential expression (Kharchenko et al. 2014; Finak et al. 2015).

Zero-inflation can also have biological origins. It has been proposed that cells produce RNA in bursts (Figure 4g) and the kinetics of polymerase binding/releasing has been fitted (Kim and Marioni 2013). We have noticed that the kinetics depends on the promoter type and that burstiness is higher for immune genes (Hagai et al. 2018). Others have seen that broadly, enhancers control burst frequency, while core promoters control burst size (Larsson et al. 2019). More work is needed to make good statistical use of what we know so far, and one must be aware that *p*-values produced by common differential expression software are greatly inflated;

and some genes fit the statistical distributions less than others. The need for biological replicates to obtain correct *p*-values has been raised (Squair et al. 2021). As sequencing and library preparation prices have dropped, and multiplexing has become easier, biological replicates must again become the norm. Statistically, *p*-values can be calculated over pseudobulk samples, where the counts for a gene are taken over several similar (clustered) cells, and the resulting pseudobulks are compared using bulk RNA-seq tools (Robinson et al. 2010; Love et al. 2014). This is a straightforward, albeit arbitrary method compared to a hierarchical statistical single-cell model, e.g.:

$$Counts \quad NB[\mu,\alpha], \text{ with } \mu \quad LogNormal(...)$$

A model that explicitly models each step of the biology and library preparation process can better integrate knowledge about, e.g., promoter/enhancer architecture, and make differential expression more about a specific biological aspect (promoter binding, enhancer use). As more data becomes available, and computing power increases, more elaborate statistical models can be expected.

## Statistics for ATAC-seq

ATAC-seq is fundamentally different from RNA-seq, and while there is a standard workflow (Baek and Lee 2020), the statistics have been much less discussed. Because there is no fragmentation, there is less concern about zero inflation. The counts have a rather firm upper limit, but regular Poisson statistics still fit the data well; however, some authors have preferred to binarize the data, resulting in a binary statistical distribution (e.g. done by Signac (Stuart et al. 2021)).

The open questions about ATAC-seq are rather about the meaning of the data. Tn5 has been shown to have a sequence bias and yields different data than the older DNase hypersensitivity assay (Karabacak Calviello et al. 2019). It is also possible to pinpoint transcription factor binding sites (or DNA binding proteins in general) as “holes” in the ATAC-seq peaks (Figure 4h). In comparison to ChIP-seq data, we have also seen sites with strong ChIP-seq peaks but no corresponding ATAC-seq peak (Henriksson et al. 2019), showing that Tn5 cannot always access TF sites (Figure 4i). One should thus ask what a “site” is. Enhancers are still defined using bulk ATAC-seq peak detection methods (Zhang et al. 2008; Granja et al. 2021; Stuart et al. 2021), applied to the single-cell data but ignoring which cell each fragment originates from. This simplifies analysis but likely misses out on discoveries the single-cell ATAC-seq data yet has to offer. Thus, much more work remains in the area of scATAC-seq analysis.

## Size factors and sequencing depth

Cells may differ in terms of the depth to which they are sequenced (i.e., the number of molecules counted). The reasons for this are unclear, but can be affected by inefficient cell lysis, unevenness in droplet size and content, stochastic enzymatic effect and stochasticity of sequencing. What is clear is that if cells are compared using any Euclidean-type measure, then they will be organized according to the total amount of molecules rather than which molecules are present. All packages thus normalize cells using a correcting factor, termed a “size factor” - in the simplest case, it is simply a division by the total number of molecules. More elaborate corrections are used for bulk RNA-seq (Love et al. 2014), but they take more time to compute and appear to be unnecessary for single-cell data.

Not much attention is given to the size factor, but some statistical notes are in order. First, RNA abundance differs between cell types, e.g., activated T cells can contain over ten times more RNA than naive T cells. Such differences are currently normalized away. Second, the less RNA/DNA that stems from the cell, the more reads might stem from background free RNA/DNA (Young and Behjati 2020). Dividing by the total number of molecules per cell is thus not a correct normalization for cells of low abundance. A quick solution is to remove such cells; however, because there is no clear cut-off for what constitutes a “low abundance cell”, this is not a perfect solution. Overall, the current size factor normalization appears to work well in practice but analysts must be aware of potential exceptions when this ceases to be the case.

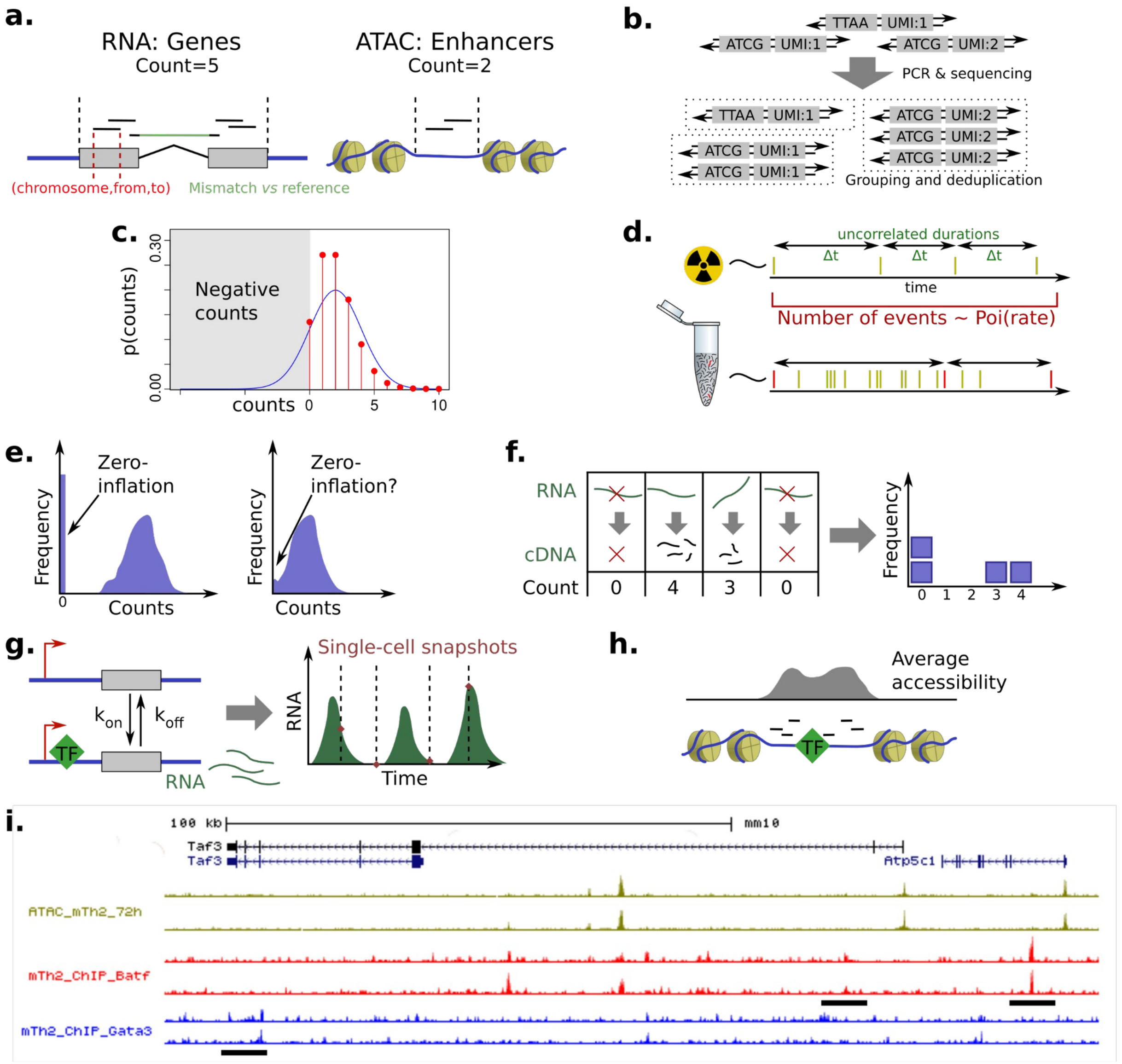

**Figure 4: From single-cell chemistry to statistics. (a)** The number of reads overlapping a feature (gene or accessible region) are summarized as a single count value. This representation is much simpler but loses information about read alignment coordinates (red) and alignment mismatches (green). For RNA-seq, gaps (green) are expected due to introns. **(b)** Molecules can be deduplicated based on UMIs; if the sequence is the same, including the UMI, then they must have the same origin. **(c)** The continuous normal distribution (blue) is not appropriate for discrete count data, such as for sequencing reads. Instead, the discrete Poisson distribution is a common choice (red). **(d)** The number of events from a memoryless radiative process, during a certain time, is Poisson distributed. Sequencing is analogous to a radiative process when the number of molecules is large, and thus the chance of picking copies of the same original DNA/RNA molecule is then low. The count for a gene (red) are then ~ Poi (TotalMoleculesSequenced * FractionBelongingToGene). **(e)** A zero-inflated distribution (top). If the amount of zero-inflation is low then it can be hard to fit (bottom). **(f)** Full-length RNA-seq gives rise to multiple cDNA fragments per RNA molecule, causing complex correlation between the counts, or alternatively modeled, zero-inflation. **(g)** RNA

might be produced in bursts, which can be modeled as polymerase on-off kinetics. **(h)** ATAC-seq data also has information about the precise binding site locations of TFs, and possibly other aspects of TF activity yet to be understood. **(i)** Example pileups of bulk ChIP-seq and ATAC-seq, where some binding sites as measured by ChIP-seq (highlighted) have no corresponding peaks in ATAC-seq data.

## From statistics to the underlying dimensions

Even for something as simple as the Normal distribution N (μ,σ), we assume that the parameters μ and σ have some sort of interpretable meaning (i.e. here, mean and variance). While single-cell data can be described by a distribution, which has over 20 000 dimensions (genes or enhancers), understanding the data means that we can find a significantly smaller set of "hyper parameters", making up a latent space to which we can assign meaning. This can be expressed as:

$$Count_i \;\; N[\mu_i(X), \sigma_i(X)], \text{ with } X \;\; SimpleSmallDistribution(...)$$

Finding a suitable latent space, and a transformation from this small space to the larger data space, is called dimensional reduction (DR). Because there is no best way of doing this, and there are tradeoffs in latent space size and shape *vs* interpretability, a plethora of approaches has been developed.

### Linear dimensional reductions

The simplest form of dimension reduction is linear DR. All linear transformations from latent space *X* to data *Y* can be described by matrices:

$$Y = WX, \vee y_{ij} = \sum_{k}^{\square} w_{ik} x_{kj}$$

Where the matrix *W* decides how the space is reshaped by combining rotations, translations and skews of the data (Figure 5a). For DR, the aim is to choose *W* such that as much information is moved to the first couple of dimensions. It is, however, not clear what, exactly, constitutes information, and the quality of the reduction relies on using the right definition.

PCA (Principal Component Analysis) is the historically most common DR algorithm (Pearson 1901) and it informed the development of all other DR methods. In PCA, the first dimensions of the latent space are called principal components, and this name is also commonly used for other methods. However, PCA itself refers to the case when *W* is chosen such as to (1) maximize the variance of the data along the first dimensions, and (2) make the dimensions uncorrelated, by being orthogonal (Figure 5b). Only the first dimensions are kept, as these are assumed to contain the largest variation and thus information. The choice of *W* is almost unique, and the computation of it is extremely fast, as it ends up being the eigenvectors of the covariance matrix. As eigenvalue problems are well studied, PCA is also rather intuitive compared to all other DR algorithms. Together, this made PCA very popular, even today, despite that the resulting DR seldom captures the relevant biology well. Ignoring the interpretability, it can also be used as a first data reduction step before using more sophisticated nonlinear DR algorithms, such as UMAP (described later). PCA can be unstable for noisy data, but improved variants exist (such as https://github.com/facebookarchive/fbpca).

Independent Component Analysis (ICA) is similar to PCA, but *W* is picked based on other criteria (Alaa 2020). Several possible criteria exist, but the intuition as to why PCA might not give the most informative latent space is shown in Figure 5c. Instead of maximizing variance, ICA can aim to maximize the skew of the data. The issue with ICA over PCA is that the answer is less constrained (not unique), and the algorithm is considerably slower. ICA has been used for single-cell analysis in, for example, the Monocle2 package (Van den Berge et al. 2020). It stands as a good reminder that PCA is not the only option.

Non-negative matrix factorization (NMF) has also been used for single-cell analysis (DeBruine et al. 2021). Unlike PCA and ICA, the requested number of reduced dimensions (here called factors) are given up front (Figure 5d). Several different numbers of dimensions are tested to find the optimal DR, but it can also be based on the expected meaning of the dimensions. Frameworks such as f-scLVM enables the analyst to also predefine some of the dimensions based on known genes (Buettner et al. 2017). This type of “bias” can help steer the model to increase interpretability. Linear models are straight-forward to extend to perform multiomics data integration (such as the MOFA package (Argelaguet et al. 2018)).

Overall, linear DR models are much easier to interpret than nonlinear models, and can easily be solved for advanced statistical distributions (ZINB and beyond). It is thus unlikely that they ever will go completely out of fashion, and even if a nonlinear model is deployed, it is still good to have a linear model to benchmark against.

## Correcting for batch effects

When single-cell libraries are generated, common variations may be introduced that affect all cells. The sources are not well understood but arise as the mixes of enzymes and buffers differ between runs. This effect (commonly known as batch effect) may cause cells to not be directly comparable. Several dedicated algorithms exist to try and correct for this effect, for example Harmony (Korsunsky et al. 2019), MNN(Haghverdi et al. 2018) and BBKNN (Polański et al. 2020). Scanpy and Seurat have other algorithms included. The problem was studied already for bulk RNA-seq data, and some older algorithms can also be used (Risso et al. 2014) given that they are fast enough for today's huge single-cell datasets. The performance of many algorithms have been benchmarked (Tran et al. 2020).

The earliest method for removing batch effects was regular PCA, where any differences in principal component 1 (PC1) frequently were due to batch differences. Thus, the differences could be handled by discarding the first component. This highlights the link of batch effects to DR. The most conservative way of statistically handling them is to introduce a categorical latent variable that represents the batch. This also holds for nonlinear methods discussed later; for example, the neural network-based model SCVI (Gayoso et al. 2022) can take the batch ID as a covariate. The only caveat with this approach is that it may be too conservative for practical use.

Modern batch correction algorithms can instead use anchor cells, which are cells deemed similar enough between datasets that they can be overlapped in the batch integration. Cosine distance has been suggested as a way of finding anchor cells (Haghverdi et al. 2018). The best way of handling batches has not yet been settled, but the idea that batch effects can correspond to a latent variable is crucial. If other experimental variables are known between single-cell datasets, it may be reasonable to expect them to appear as latent variables (implicitly or explicitly).

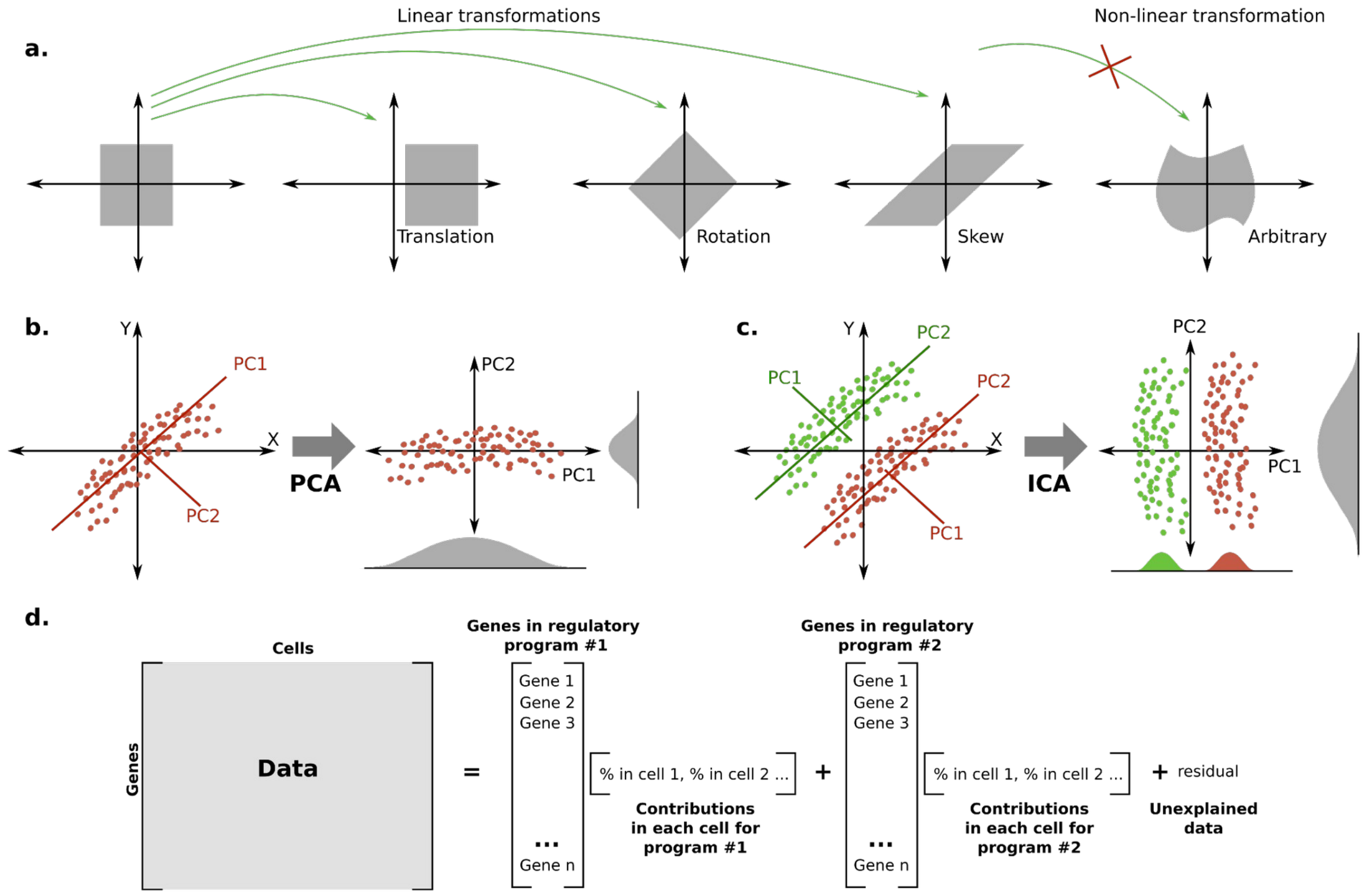

**Figure 5: From statistics to the underlying dimensions. (a)** Linear transformations are limited to combinations of translations, rotations and skews. **(b)** PCA finds a new coordinate system that maximizes variance along the first dimensions. **(c)** ICA separates two populations of cells along PC1, while PCA rather would put the component with highest variance as PC1 (here PC2). **(d)** NMF performs a linear decomposition, with the number of latent dimensions predefined. In advanced factor analysis models, the content of the dimensions can also be partially specified.

# Cell states and latent space topology

nonlinear models were developed because many processes (including many in biology) simply aren't linear. Unfortunately many nonlinear models fit to the same data, and they are hard to interpret (Figure 6a). Before using a nonlinear model, it is thus important to have an idea of how one wants the model to behave, and how it might behave. This requires a fair bit of abstract mathematical thinking that will be presented here.

Topology is a subdiscipline of mathematics that is focused on the properties of spaces and surfaces(Armstrong). It tries to make concepts concrete, such as path-connectedness. For example, one might ask the question, “is there a way to connect a point A to a point B?”. If point A is taken to be among pluripotent cells, and B is taken to another cell type, this type of connectedness is equivalent to the biological question ”can type A cells differentiate into type B cells?” (Figure 6b). While this question is fairly easy to answer by a human if the cells are reduced to a 2D plane, it is not obvious in the higher dimensional (20 000 for all genes) space. Luckily, the field of topology is set up to handle any number of dimensions. Anyone analyzing higher dimensional data should thus be interested in topology.

## Discrete topologies and basic nonlinear dimensional reduction

One problem with topology is that mathematicians usually have well-defined spaces (based on an equation). Single-cell biologists, on the other hand, just have a finite set of noisy observations. A link is achieved by approximating topologies from the data points using k-simplexes - a multidimensional equivalent to a triangle (Figure 6c). These simplexes can be built by letting each corner be a single-cell observation. The edges remain lines, and the lines can be picked by distance. How they should be picked is an open problem, but typically cells are connected to the k nearest other cells, also known as the nearest neighbors (the result is called a kNN-graph, Figure 6d). If an Euclidean space is assumed, i.e. the Euclidean distance is used, then the kNN-graph can be computed quickly also for large numbers of cells.

The kNN-graph is the input to the most commonly used nonlinear DR tools for single-cell data: t-distributed stochastic neighbor embedding (t-SNE) and uniform manifold approximation and projection (UMAP) (McInnes et al. 2018). While the algorithms can work with the full set of cell-cell distances, the closest neighbors are the most relevant, and focusing on these speeds up the computation tremendously. The output of each of these algorithms are 2D coordinates (or user choice of dimensions, with 3D frequently being useful), where the distance of the points reflects their distance in the higher dimensional space (Figure 6e).

The greatest issue with t-SNE and UMAP is that interpretation is difficult (discussed further in the later section on cell types). While these methods are good to give an unbiased overview of the data, the latent space axes have no meaning, and the distance might not reflect the most interesting biology. Thus, other nonlinear models that help capture the biology of interest are recommended after initial data investigation.

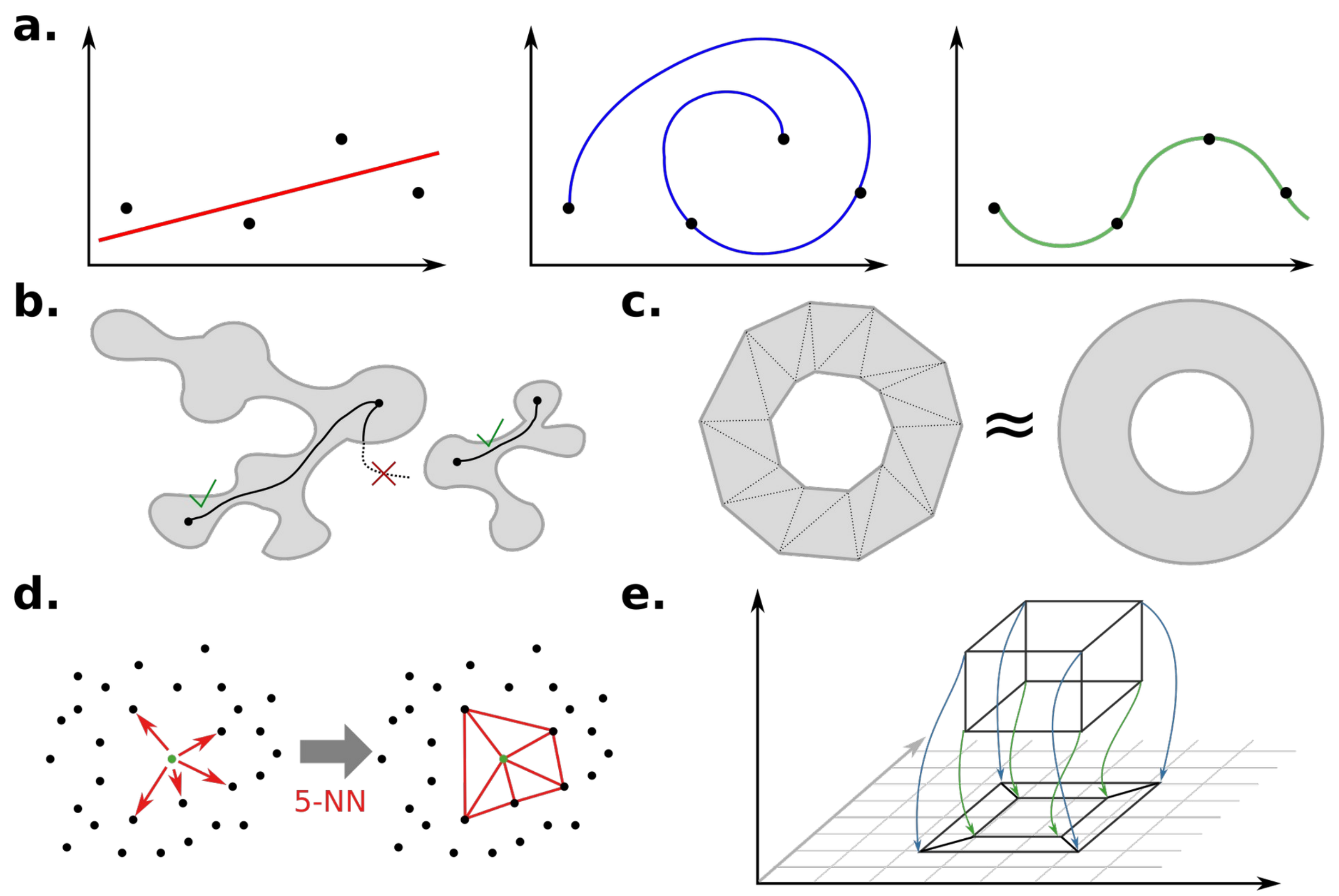

**Figure 6: Cell states and latent space topology. (a)** A linear model (red) has a single solution and easy interpretation while nonlinear models (blue, green) fit the data better but are hard to interpret. **(b)** Whether a cell type can differentiate into another cell type can be seen as a topological problem of connectedness; this particular topology has the shape of a tree. **(c)** Spaces or topologies can be approximated arbitrarily well by triangles; or simplexes in higher dimensions. **(d)** The 5 nearest neighbors can make up a local graph or local simplex. **(e)** UMAP and t-SNE reduces the number of dimensions while trying to retain distances, effectively retaining topology as well as possible. The optimal solution is, however, not unique.

## Cell state space dynamics and trajectory inference

While t-SNE and UMAP aim to simplify the data in terms of mapping them directly to a lower dimensional space, another approach is to find a simplified topology. These topologies can, but need not, lend themselves to easy presentation. While nonlinear factor analysis can handle topologies such as lines and planes, it cannot handle tree topologies (Figure 6b), as relevant for cell fate decisions during, for example, differentiation. Because of the historically close link to differentiation over time, the use of algorithms to analyze line or tree topologies are called pseudotime analysis, or trajectory inference. Several pseudotime algorithms have been proposed and compared (Saelens et al. 2019).

Trajectory inference rests on several assumptions. First, if the trajectory inference is for a time-based process, then the data must contain cells representing all the time points. Since cells commonly do not respond at the same rate, this is frequently the case, but cells from several time points may need to be mixed. Most crucially, no algorithm can prove that a pseudotime trajectory exists - it is an assumption (RNA velocity (Bergen et al. 2021), and RNA metabolic labeling (Qiu et al. 2022), which measures the vector field, tries to overcome this limitation). Some algorithms try to find the type of trajectories, while others simply accept user input. Thus, the latent space can be shaped by prior knowledge, or it can be unbiased.

Several algorithms are based on algorithms that find the Minimum Spanning Trees (MST), i.e. the smallest subset of edges in a graph, given weights (distances), that still connect all the vertices (Figure 7a). For single-cell data, the kNN-graph is the input to the MST algorithm. However, due to noise, the MST graph can become rather complicated. To avoid overfitting, several methods are thus used to simplify the graph. Slingshot (Street et al. 2018) and Monocle(Trapnell et al. 2014) are examples of MST-based algorithms. MST has the advantage (and disadvantage) of not enforcing the number of end-points *per se*. Other approaches exist that simplify the graph, not necessarily to a tree, such as PAGA (Wolf et al. 2019).

Knowledge can be extracted from the graph representations in several ways. For simple graphs, such as trees, a subset of cells can be ordered from the tip of one branch to another. Another approach is to study the dynamics of cells, assuming they transition semi-randomly over the neighbor graph. The simplest suitable statistical models are memory-less processes. These also correspond to a Newtonian model of the evolution of the cell state *X* at time t, which is assumed to contain all information needed to predict the future:

$$dX/dt = f(x) + noise \Leftrightarrow p(X_{t+1} \vee X_t) = f(X_t)$$

Memory-less processes of this kind are also called Markov chains, and can be described by graphs having transition probabilities on each edge (Norris 1997) (Figure 7b). Thus, the single-cell neighbor graph with suitable probabilities assigned can be treated with powerful Markov chain theory. The jumping probabilities can be uniform, or informed by other data such as RNA velocity (Lange et al. 2022). If some vertices in the graph only have incoming edges, then these are denoted as *absorbing states* - a random walk will at some point get stuck in any of these (Figure 7c). If there are multiple such absorbing states then it is easy to compute which end-state is the most likely, and what is the average number of random jumps until it happens (corresponding to total time). This can be used to estimate the likelihood of a type differentiation. If no absorbing states exist, then it is possible to calculate the *stationary distribution* - how likely it is for a cell to be in a given state, independent of where the cell starts (assuming a property called *ergodicity*, usually fulfilled for single-cell data). These are just the simplest markov chain concepts, and the chain can be designed to model and answer various biological questions.

One caveat with current pseudotime methods is that they provide no information about *why* any branching occurs. An analogy is with a previous study of speciation of ours(Henriksson et al. 2010): to learn why speciation occurs, the concept of speciation had to be removed from the model. Similarly, it may be speculated, cell fate decision making cannot be understood using algorithms that enforce a simplified model of a branching event. In particular, the local topology at the branch is not understood at all. Unbiased Markov chain-based modeling is likely a good bet for anyone trying to further understand cell fates.

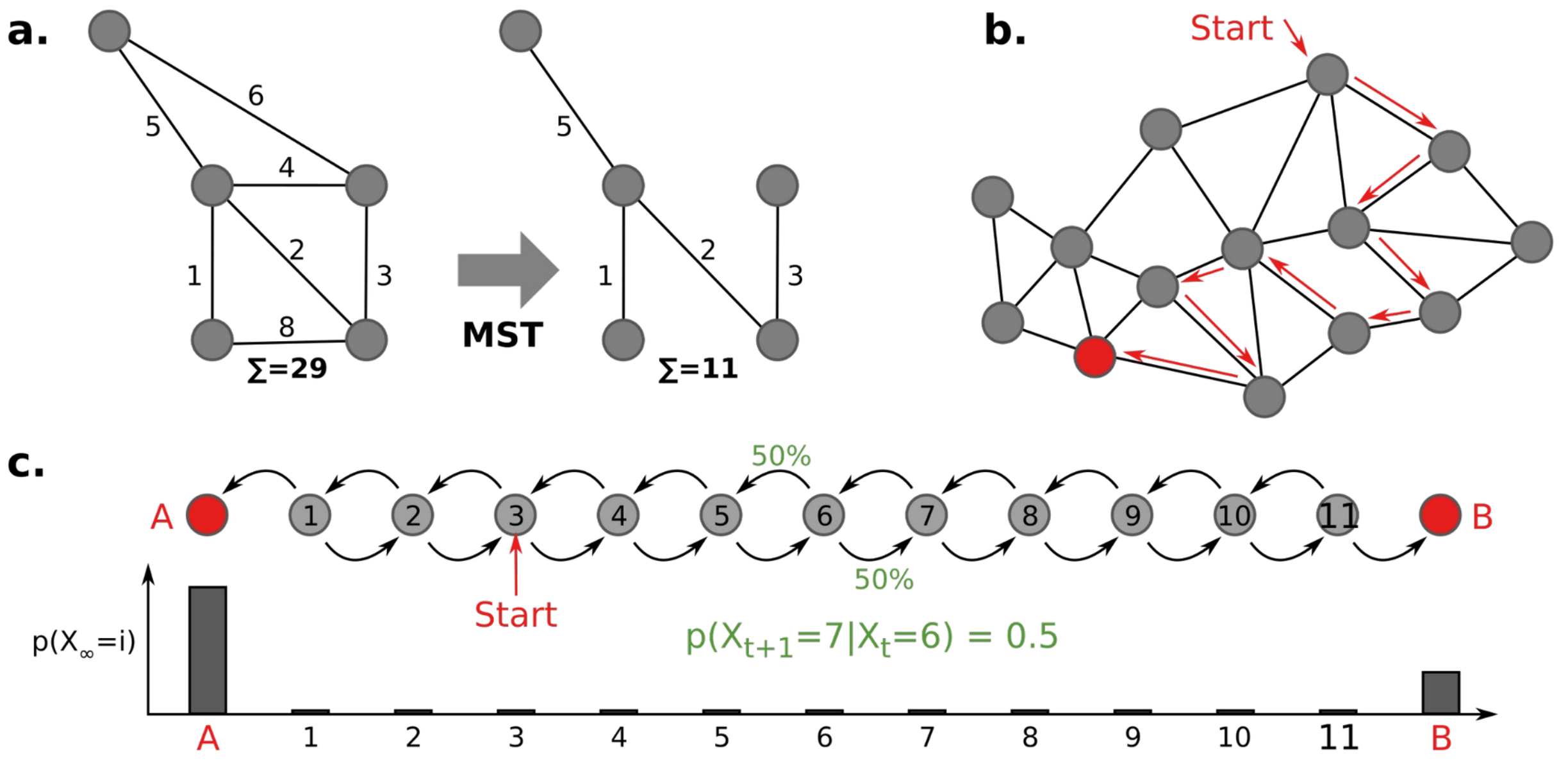

**Figure 7. Cell state space dynamics and trajectory inference**.
**(a)** The Minimal Spanning Tree (MST) of a graph can be used as a simplification step. **(b)** Cells can be thought to move randomly over the graph, representing progress (differentiation, activation) in cell state space. **(c)** If there are only incoming edges to a node, from no outgoing edges, then the node is called an absorbing state. Given a start position, the probability of ending up in either A or B can be calculated. In this example, the probability of absorption in A is higher than in B, assuming equal jumping probabilities.

## Donuts and the cell cycle

Topology offers tools to reason about the shape of the latent space. One of the most famous results from topology is that a coffee cup is equivalent to a donut (Figure 8a). This result more specifically tells us about a certain type of connectedness: can one line from A to B be deformed in a continuous manner such as to overlap another line from A to B? It turns out that there are lines on the donut which do not have this property (Figure 8b), because the hole in the middle restricts the deformations. The same property holds for a coffee cup. In topology, the coffee cup is equivalent to the donut in the sense of having similar behavior line-deformation-connectedness properties.

The concept of topological equivalence is important to us because we can reason about simpler spaces instead of the high-dimensional raw data. The cell cycle can be thought of as a circle (which indeed is the way it is commonly drawn, Figure 8c). The cell cycle state can be identified from RNA-seq data using common workflows (part of both Seurat and SCANPY), and cells can be annotated as being in either G1, S or G2M phase. However, the standard workflow does not *order* the cells within these phases, limiting the resolution at which cell-cycle linked events can be studied. In limited cases, it might be possible to use a pseudotime algorithm, but because *linear* pseudotime has the implicit assumption of a start and an end, *linear* pseudotime is topologically incompatible with the cell *cycle* (Figure 8c).

There are several attempts at ordering the cells according to the cell cycle. The common RNA-seq analysis method for categorizing cell cycle state is based on a list of marker genes, and a PCA is used to reduce the number of dimensions to 2. It is likely not a coincidence that the smallest number of dimensions in which a circle (or cycle) can be represented is also 2. In principle, the angle in this reduced space can be used to order the cells, although we have never seen this performed.

One caveat with topological reasoning is that we only have a finite number of samples, and thus only approximate knowledge of the space. Can cells ever reside in holes of the topology? If points or the kNN-graph is assumed to have volume, then this question can be addressed by testing different volumes. This has been applied to single-cell data (Rizvi et al. 2017) and the concept enables a range of analyses, such as how "small" or isolated a gene regulatory program is (Figure 8d).

## Housekeeping genes, group theory, and product spaces

Group theory is a topic of abstract algebra and describes the mathematical structures (groups) generated by binary operators, such as addition or multiplication. A formal definition is beyond this review, but the main use is in analyzing symmetries. It has had success in X-ray crystallography, where it can be used to prove the number of possible crystal structure symmetries. While addition is a function over the space of numbers, it can also be used to organize, e.g., how many turns plasmid dsDNA is wounded. The state can be represented by the winding number (Figure 8e). The operator of interest is in this case "PositiveWinding", though an inverse "NegativeWinding" can be derived. This winding operator operates on circular dsDNA as objects, adding the turns of one plasmid to another. Because winding behaves exactly as integers over addition, an isomorphism can be defined over *classes of topologies* to *numbers* (thus, one can speak of winding numbers, instead of complex geometric objects). The main point of group theory is that various symmetries can be argued based on the properties of the operator, and the operator can operate over topologies. Conversely, symmetry usually implies some type of group, e.g., a rotated circle is still a circle (making the rotation operator an identity operator in this case). The breaking of symmetries has also been studied in relation to groups in embryogenesis (Kumar and Bentley 2003). Past positive examples make use of group theory tempting also in a single-cell latent space setting.

Groups are related to product spaces. We can imagine a representation of the cell state to be a position in "the space of cell cycle", and simultaneously a position in "the space of cell differentiation". The set of all possible two positions together (cell cycle, differentiation) then make up a higher dimensional "product space" (Figure 8f). Such a latent space can naturally be expected to capture more of the biology than just each of them alone.

Unfortunately, there are few pure "operators" in biology that operate on only one space, and thus the product space analogy breaks down. Even processes considered to be "housekeeping" are intertwined with other processes. Several examples can be given for the cell cycle: activation of naive T cells is essentially a synchronized entry into the cell cycle; and cells in the skin preferentially divide during the night (Beri and Milgraum 2016), linking it with circadian rhythm. T cell migration in and out of lymph nodes is also linked with the circadian rhythm (Druzd et al. 2017). These are just some examples of how seemingly housekeeping processes are linked to more specialized processes, and how there are no truly independent spaces. This is the most important take home-message - biology doesn't conform well to idealized representations such as independent spaces. Nevertheless, in limited cases, it can be a good approximation, and fitting data to idealized models is a natural part of hypothesis testing.

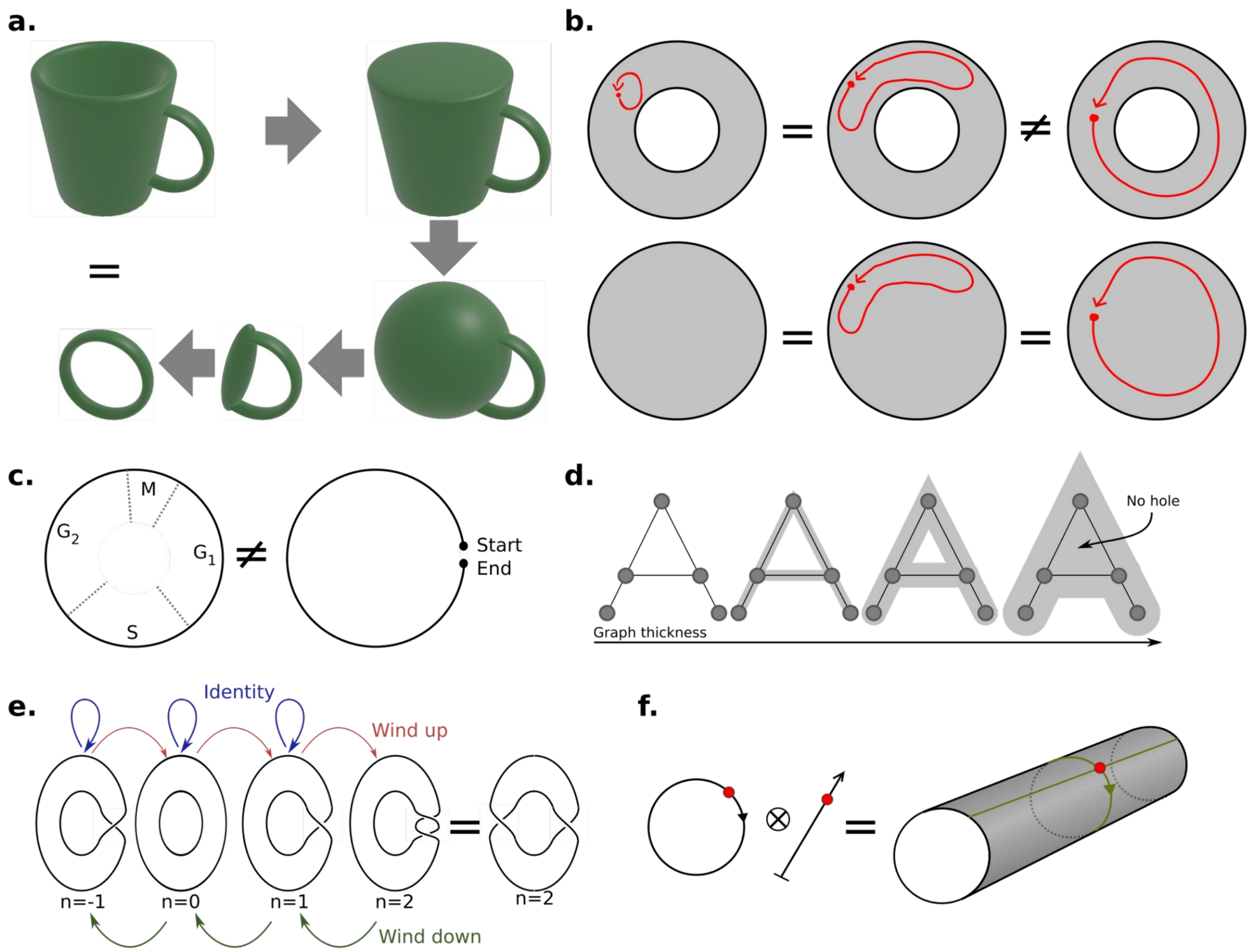

**Figure 8. Topological reasoning. (a)** A coffee cup and a donut are topologically equivalent based on their connectedness properties, and can further be continuously deformed into each other. **(b)** Lines on a donut cannot always be continuously deformed into each other because of the hole in the middle. But if the hole is removed, the topology becomes path-connected. Thus a circle and open plane are not topologically equivalent. **(c)** The cell cycle has the topology of a circle, which is different from a line. **(d)** If the graph is given thickness, then the existence of topological holes depends on the thickness. **(e)** The winding number can be used to describe how twisted plasmid dsDNA is, and organizes different conformations into equivalence groups. It is generated from up- and down-winding operators. **(f)** The product space generated from combining the cell cycle (circular) and differentiation level (from low to high, linear) is a cylinder. In reality, a cylinder latent space might however not fit the data.

## Nonlinear models and neural networks

Graph-based reasoning allows one to compare cells in a topologically relevant manner, but it does not provide an explicit function to and from the latent space to the data space the way PCA, NMF, or other methods do. Having an explicit function is a requirement for using most statistical tools. Unfortunately, it is normally difficult to motivate the choice of a nonlinear function for higher-dimensional data. One way out is to allow a large range of differently shaped functions; this can be done by using neural networks (NNs). These are inspired by neurons, built up by many small simple units that together can produce complex behavior. The smallest modern "neuron" typically looks like this (Figure 9a,b):

$$Y = ReLU \text{¿}$$

Where the nonlinear function ReLU is defined as follows:

$$ReLU(x) = \{0\, if\; x < 0 ;\, x\, if\; X \geq 0\}$$

Several layers of neurons make up a neural network (Figure 9c). The optimal input weights $w_i$ and b are computed using optimization, e.g. by minimizing the difference of the neural network output *vs* the given data (also known as the reconstruction error):

$$Minar\, g_w \sum_i^{\square} \left| Y_i - NN_w(X_i) \right|^2$$

There is little special with this mathematical construction, except that it has a "well-behaved" differential, which helps optimization over many layers of neurons (avoiding what is called the "vanishing gradient" problem, which occurs when lower-layer connection weights in a deep NN become static, hindering or halting further training of the NN). Secondly, it is in fact just many multiplications and additions organized in a coherent way, and computing it fits well with how graphics processing units (GPUs, graphics cards) are designed. GPUs enable an order of magnitude faster solving of NNs. Since NNs are nowadays easy to use and fast to compute, many methods are based on them. Note, however, that linear functions are a special case of neural networks, and everything here could be designed for these as well.

A final important note is that the NN reconstruction does not rely on using the Euclidean distance between data points. The reconstruction error model can easily be modified to better incorporate statistical properties. Because the Euclidean distance enforces a certain type of topology, and thus latent space structure, the use of NNs can thus open the door to biologically more relevant latent spaces.

### Generative processes

Methods such as PCA, ICA, NMF, and UMAP, were primarily developed to map points *from* data *to* latent space. An alternative approach is to develop algorithms that map *from* latent space to *data*. However, since the latent space has a smaller dimension, it cannot easily cover all of the larger space (Figure 9d; filling the data space is possible with space-filling curves (Armstrong), but they are only of theoretical interest). A solution to this is to consider multi-valued functions; that is, f (*x*) can return more than one value. In the case of *generative processes (GPs),* the returned values follow a probability distribution. A basic generative process from latent space *X* to data space *Y* can look as follows (Figure 9e):

$$x \quad Uniform[0,20]$$

$$y_1 \quad Poisson[\lambda = x]$$

$$y_2 \quad Normal[\mu = 0 \, if \, x < 10, otherwise \, 5 ; \sigma = 1]$$

This generative process was constructed manually and illustrates the concept. It shows the enormous freedom in the choice of latent space and GP. However, for most applications, an algorithm is used to fit a suitable GP to the data. Because these algorithms have little *a priori* knowledge of the data, highly flexible NN nonlinear functions are commonly used in combination with simple statistical distributions. A hypothetical example could be:

$y_{i,j} \quad Poisson[\lambda = NN_j(x_{i,1} \ldots x_{i,m})]$, i over all cells and j over all genes

Quasi-linear versions have also been tested on single-cell data, with the aim of providing explainable mappings (Svensson et al. 2020). To find the best NN, this has to be recast as an optimization problem. Two examples of networks/algorithms that realize this in practice are Variational autoencoders (VAEs) and Generative Adversarial Networks (GANs).

## Autoencoders (AEs) and Variational Autoencoders (VAEs)

Autoencoders (AEs) were first developed independent of GPs, but were later fused with the concept, resulting in Variational Autoencoders (VAEs). In addition to the function from latent space to data (in this context, the GP is called the *decoder*), a function is also sought from data to the latent space (called the *encoder*). These are considered meaningful, if given a data point, it can be encoded, and then decoded, the data point approximately reconstructed (Fig 9f). The solution to the reconstruction problem is typically not unique, giving rise to many possible latent spaces. Furthermore, it might not from a human standpoint have sufficient structure (Fig 9g).

To enforce a meaningful latent space structure, plain AEs have been replaced by VAEs. In the simplest description, the idea is that samples close in the latent space should correspond to similar output data. Thus the latent space is smoothened, avoiding the scenario in Fig 9e. However, to gain a deeper understanding and appreciation of how they work requires a proper mathematical treatment.

For VAEs, unlike AEs, the latent space is a probability distribution (Figure 9h). This distribution is commonly called the variational distribution, Q (*z*|*x*). The latent variables, *z*, are sampled from this distribution, and the random sampling is what ensures that similar latent space points result in similar data points. Figure 9f shows the basic structure of a VAE, where the NN parameters $\theta$ and $\phi$ are found by minimizing the following loss function over all data points $x_i$:

$$l_i(\theta, \phi) = E_{z \sim q_\theta(z \vee x_i)}[\log p_\phi(x_i \vee z)] - KL(q_\theta(z \vee x_i) \vee ¿ \, p(z))$$

The first part of the function is the reconstruction loss, going from latent space *z* to data space *x*. In other words, it is a measurement for how similar the reconstructed data is to the original input. The second half of the loss function measures the KL-divergence (a measure of difference between two distributions) between the variational distribution and $p(z)$. The distribution of $p(z)$ greatly determines the final latent space structure but is almost always set to be a Normal distribution, a rather unbiased choice.

The minimization of the loss function could be done with any nonlinear optimizer, but because of the random sampling of the latent variables $z$ from the variational distribution $q_\theta(z \vee x_i)$, convergence would be poor. The solution to this problem is known as the reparameterization trick. Unfortunately this trick limits which variational distributions can be used for the latent space, as the distribution must have certain symmetries. The Gaussian distribution, which is commonly used, is sufficient for most cases; however, other distributions have been tested (Ding and Regev 2021).

## VAEs for single-cell analysis

The link between VAEs and the known single-cell statistics is through the GP $p_\phi(x_i \vee z)$. In other words, given a latent space point, what is the statistical distribution of the data? The toolkit SCVI (Gayoso et al. 2022) proposes the following GP for RNA-seq (Lopez et al. 2018) (simplified to show key concepts). The latent space has been split into *z* and *L*, and below is thus actually $p_\phi(x_i \vee z, L)$:

$z \;\; Normal[\mu=0, \sigma=1]$

$L \;\; LogNormal[\mu, \sigma]$ with $\sigma$ estimated from data

$\varrho \;\; NN(z, covariates)$

$X \;\; ZINB(\mu=\varrho L, dispersion=NN(z))$

This real life example shows how different types of data distributions can be modeled on top of a latent space. The latent space has been split such that *z* represents cell type, and *L* the sequencing depth (size factor). The log normal distribution of *L* ensures that the size factor is always positive and close to the fitted value. However, the model also works well if *L* is just replaced with the fitted size factor for the particular cell. Next, the variable $\varrho$ is found using a NN, which optionally can use information about e.g. which batch the cell comes from (for batch correction). The variable $\varrho$ can be thought of as to represent the RNA levels of an ideal cell, without zero inflation, or differences in sequencing depth. To account for sequencing depth, it is later simply rescaled as $\varrho$*L. Finally, the data *X* follows a ZINB-distribution with $\varrho$*L as the idealized average cell. However, NB or Poisson can equally well be used instead.

This example shows the enormous flexibility of the VAE framework. Given a latent space, which can be given any number of dimensions and shape, a neural network can transform it into the parameters of any choice of probability distribution. Other GPs represent CITE-seq (Gayoso et al. 2021) and ATAC-seq (Xiong et al. 2019; Ashuach et al. 2022), and can easily share the latent space z for multiomics integration.

Several open problems remain for VAEs. In SCVI, a plain neural network is used to encode data into *z*, without any use of known single-cell statistics, and can thus limit the reconstruction. Also, several users have reported VAEs to be "brittle", giving rather different solutions for small changes in the input, or changes in the NN structure. If the data is limited then, as for most NN-based algorithms, VAEs struggle to fit a good model. This requires tuning of the NN architecture, in terms of number of layers and neurons in each layer. An ideal model of data should not rely on technical parameters, and in this regard, VAEs have a long road left ahead. Nevertheless, their flexibility in statistical formulation suggests that the trip is worth the effort.

## Graph neural networks

A class of neural network-based algorithms operate on data organized in graphs: Graph Convolutional Networks (GCNs). They have for example been used to predict properties of molecules, where the atoms and their connectivities make up a graph (Reiser et al. 2022). The famous AlphaFold algorithm for predicting protein structures is also a GCN (Jumper et al. 2021). For single-cell data, the graph is commonly, but not necessarily, the kNN-graph. While GCNs thus rely on the latent space induced by the Euclidean distance metric, they permit entirely different problem formulations.

For GCNs, the question is: can knowledge be gathered at a local point of the graph, NN ($G_{local}$), and can it be sufficiently propagated by repeated application of NN layers, $NN_1$ ($NN_2$ ($G_{local}$)) (Figure 9i)? This approach can be motivated by a field of mathematics, fixed point theory, studying functions and points $x_{fix}$ such that $f(x_{fix})=x_{fix}$. This is related to the repeated application of functions, and convergence to these fixed points: $f(f(f(\ldots f(x))))\rightarrow x_{fix}$. Such a fixed point would represent the fully extracted knowledge of the graph.

Because GCNs are such a broad topic, and further can be combined with VAEs (Wang et al. 2021), this section primarily lists some useful single-cell applications. Here ~ should be read as some sort of VAE formulation, and the annotation is primarily for illustration:

- **GeneExpression = F (G).** In this case, the aim is to compute the expression levels of this cell by comparing it to the neighboring cells. This can be used to calculate the expression of an "idealized cell", without the technical or biological noise - also called denoising. Some downstream algorithms prefer smoothened data, and it can also be used to aid visualization.
- **CellType = F (G)**. Instead of relying on clustering, and annotating cell types from the average gene expression levels of that cluster, it can be done directly by investigating each cell and the neighboring cell. Since clustering requires manually providing settings about the resolution (or expected cluster size), the GCN approach is less arbitrary.
- **$F_{protein}$ (G) ~ $F_{ATAC}$ (G) ~ $F_{RNA}$ (G)**. Different data modalities can be compared (or "integrated") using some variant of graph neural networks. Again, this avoids the need for clustering, but it further has the advantage that the graphs need not be the same. For example, the graphs based on ATAC-seq need not correspond to those for RNA-seq.

- **$F_{RNA}$ ($G_{RNA}$) ~ $F_{gene_homology}$ ($G_{genes}$)**. There are cases when two orthogonal graphs are being estimated. One such case is the comparison of cells between species; the traditional euclidean distance between cells is problematic because it is not clear which genes in species A should be compared to which genes in species B. However, assuming that the cells are lined up correctly, and with some knowledge of homology (based on gene sequences) it is possible to find which genes correlate and thus correspond. This is a circular dependency: The cell-cell correspondences depend on the gene-gene correspondence which depends on the cell-cell correspondences. Circular problems of this type beg for an iterative algorithm that solves both problems at the same time. GCN fitting is iterative and naturally matches the structure of the problem.

It is still early days for the use of graph NNs for single-cell analysis. Likely any single-cell problem can be formulated within this framework, and the combination with VAEs also enable them to capture an interpretable latent space. GCNs are thus a prime area for novel research.

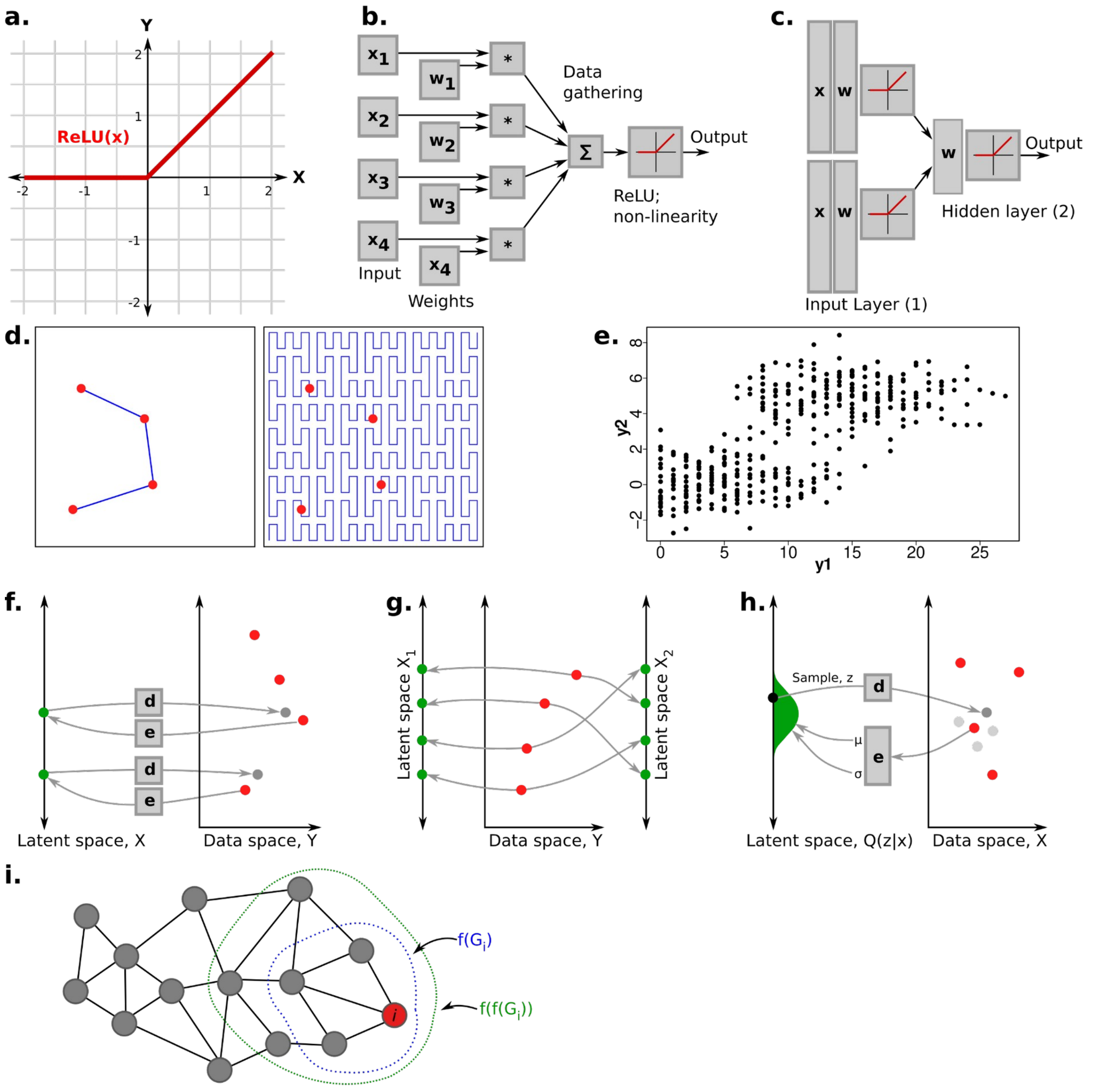

**Figure 9: Generative processes and neural networks. (a)** The ReLU function is a simple nonlinear function that forms the bases of modern neural networks. **(b)** One neural network layer, showing how weights are applied to each input value, and a sum and ReLU element turn all inputs into a single output value. **(c)** Several neural network layers are stacked to increase nonlinearity. **(d)** A mapping from a lower dimensional space cannot cover a higher dimensional space unless it is extremely contorted (space filling Peano curve on the right, source: wikipedia). Such mappings struggle to retain the natural underlying topology. **(e)** Example of the generating process: $y_1$ ~ Poisson[ $\lambda$=x ], $y_2$ ~ Normal[ $\mu$=0 if x<10, otherwise 5; $\sigma$=1 ], x~Uniform[0,20] **(f)** The basic AE encoder and decoder paradigm to reconstruct data from a lower dimensional space. **(g)** Two latent space encodings (left and right) of a higher dimensional space (middle). The solution is not unique but the left encoding captures the topological structure better. **(h)** The VAE architecture map data points to *probability distributions* in latent space Z. Here the distribution is Normal ($\mu$,$\sigma$) with the parameters

encoded by a NN (left). Random samples from this distribution are then drawn for reconstruction by another NN (right). **(i)** Graph convolutional networks (GCNs) perform repeated application of a function over local nodes of a graph, aggregating and propagating local information. Here the information gathered for two applications of the NN are shown (once in blue, twice in red).

## Clustering, language, and cell types

The most primitive latent space representation is that of categories of cells. Formally:

$Y_i \;\; SomeDistribution(f(c_i))$ for some category of cell $c_i$.

The representation has the advantage that it is easy to perform pairwise comparisons of clusters, and the categories can be given memorable names. Batches and different treatments of cells usually map to categories, but categories are primarily discussed in the context of cell types.

There are many ways of classifying cells as different types, including tools that compare the gene expression to databases of profiles (Abdelaal et al. 2019). However, the most common methods split the kNN-graph in such a way that the sum (“cost”) of the cut edges is minimized (Figure 10a). The family of spectral graph cutting methods has not gained much traction, but has been used (Schwartz et al. 2020). The by far most common methods are now Louvain (Blondel et al. 2008) and Leiden (Traag et al. 2019), available through packages like Scanpy and Seurat. The use of graph cutting clustering algorithms thus makes the matter of cell type categorization a topological one, where the algorithm objective function need not correspond to what the user wanted (Figure 10b). The number of clusters also has to be specified by the user and there is seldom an objectively “correct” number of clusters. This brings this review to the longest standing open question in the single-cell community.

### What is a cell type?

Cell types were originally identified by their *morphology* as this property was first available. The *function* of the cells also entered the definitions. Neurons and muscle cells are very distinct. This is a type of natural history, where cells were simply grouped together (like all biological samples) to create a sense of order. Overall, cell type definitions have largely followed what industry and technology has had to offer (such as microscopes), the intertwining called technoscience (Pickstone 2001). As new measurement tools became available, such as Fluorescence-activated Cell Sorting (FACS) machines to sort by surface marker proteins, definitions have increasingly moved to be based on the *cellular content*. FACS has played a major role in immunology, where a huge number of cells have been defined by an increasing list of proteins.

Sequencing and proteomics have later challenged old surface marker based definitions. One issue is that some marker proteins have only low levels of corresponding mRNA, and thus aren’t suitable for single-cell RNA-seq cell annotation. Another even bigger issue is that the mRNA need not be present in every cell, despite the protein being there. A debate has raged whether this is for technical or biological reasons, but consensus moves toward the idea that mRNA is produced in bursts. Because the protein carries the actual function, the mRNA need not be present at all times, thus pointing towards inherent biological reasons. The solution in the single-cell world has been to rely on unsupervised clustering and using the average profile for annotation. When clusters are distinct, this is usually beyond doubt. However, for some cells, with subtypes of neurons, T cells, and monocytes as notable examples, the boundaries between classically defined subtypes are not clear (Figure 10c).

A way around the unclear boundaries is to refer to the cell differentiation history, classically called the “lineage tree”. Thus, a cell type is not just a separate category, but somehow linked to other categories with shorter or longer distance. For comparison with other types of cells, it primarily makes sense to compare with other cell types in the same lineage (Figure 10d). This is already implicitly performed by analysts by setting the clustering resolution appropriately, but the lineage relationship could in principle be stored in the annotation as well. Concepts from comparison of genes across species, such as in-paralogs (Sonnhammer and Östlund 2015), could be borrowed to make the correct comparisons more formal.

However, in many single-cell datasets, a cluster exists based on the cell cycle alone. Are dividing cells their own cell type? Many would argue against it and would rather call it a *cell state*, thus questioning the ability of current clustering methods to define cell types. The solution calls for clustering algorithms that either interpret the latent space topology differently, or topologies that fit better with our notion of cell types. Alternatively, the nomenclature needs updating. The Human Cell Atlas is, for example, trying to update the cell type definitions (Osumi-Sutherland et al. 2021). But sometimes it is not even clear what should be annotated (Figure 10e); muscle cells are nucleated, suggesting that the outer cell membrane is the object of interest. However, epithelial cells are stuck together and share space through gap junctions, suggesting that epithelia is just one single large cell. This motivates a higher level view on the cell type problem.

Sociologists provide several external views. One view emphasizes rather the power relationships (Foucault 1995), such as reviewers for grant agencies and journals upholding the use of certain terminology (the poststructuralist view). Different review boards may also be interested in different genes based on technologies, e.g. for T helper type 2 cells, a genomics panel would be concerned with GATA3 expression using sequencing, while an immunologist panel would look at IL4 secretion using flow cytometry (the historical materialism / Marxist view). The constructivist view looks at this simply as new concepts being developed in the light of old concepts, and that as we specialize and socially compartmentalize, we develop different new concepts that need not agree. The logician Wittgenstein argues, using his language game model (Wittgenstein 1998), that language need not be “logically correct” - just serving a function within a certain context. Interestingly, the ML community is just about to bring back many of these abstract and questioned concepts, but in a new quantifiable shape.

## Semantics and language models

The use of language models deserves extra attention in the context of latent representations. What is knowledge and what is meaning? This philosophical problem has a long history, and with the advent of logic, it has been argued that “meaning” is what someone tries to convey in a sentence (or utterance). Sentences can then be described in a type of logic (T-theories in the Davidsonian tradition of philosophy (Glüer 2011)). This idea has caught on in the study of causality. Classical statistics does not handle causality, only correlation; it has been speculated that the reason causality has been mathematically largely undeveloped until recently, is because language is such a great causal inference system (Pearl and Mackenzie 2018). Causality is, however, no simple matter, and already Aristotele tried to understand it in more detail (Haig 2020). If we think of understanding biology, in terms of causality (“calcium is released because X bound to Y”), then this can be understood in terms of language, but we still do not understand language itself well. However, computers are good at modeling language (natural language processing, NLP); does this mean that computers can understand biology? And how well can language be a suitable latent representation for the data we see, that is (Figure 10f):

$Y_i \quad SomeDistribution(f(u_i))$ for some verbal description $u_i$ about cell i.

The NLP field is these days centered around transformers (Vaswani et al. 2017), a type of NN over strings of data that excels at keeping references to earlier parts of the data string (Figure 10g). If the data string is a set of characters or words, then it maps directly to language processing. But transformers can also process sound, and more recently it has even generated realistic images from textual descriptions(Ramesh et al. 2021). One type of transformer, BERT (Bidirectional Encoder Representations from Transformers) (Devlin et al. 2018), has already been applied to single-cell data (Yang et al. 2022).

From the discussion of clusters, and how these do not always map well to classical cell types, one point about language should already be clear: if our language is imprecise, then so is our ability to model and understand. If the vocabulary is poor, then this limits any statistical model on top. However, this field has a future in that at some point, our models must map to human understanding, which arguably is based on the brain's language center. It must also connect to previous research, written up as language in articles and largely inaccessible to anything but NLP. This is thus a topic that should be followed by great interest and especially taught in any biology class on the theory of science.

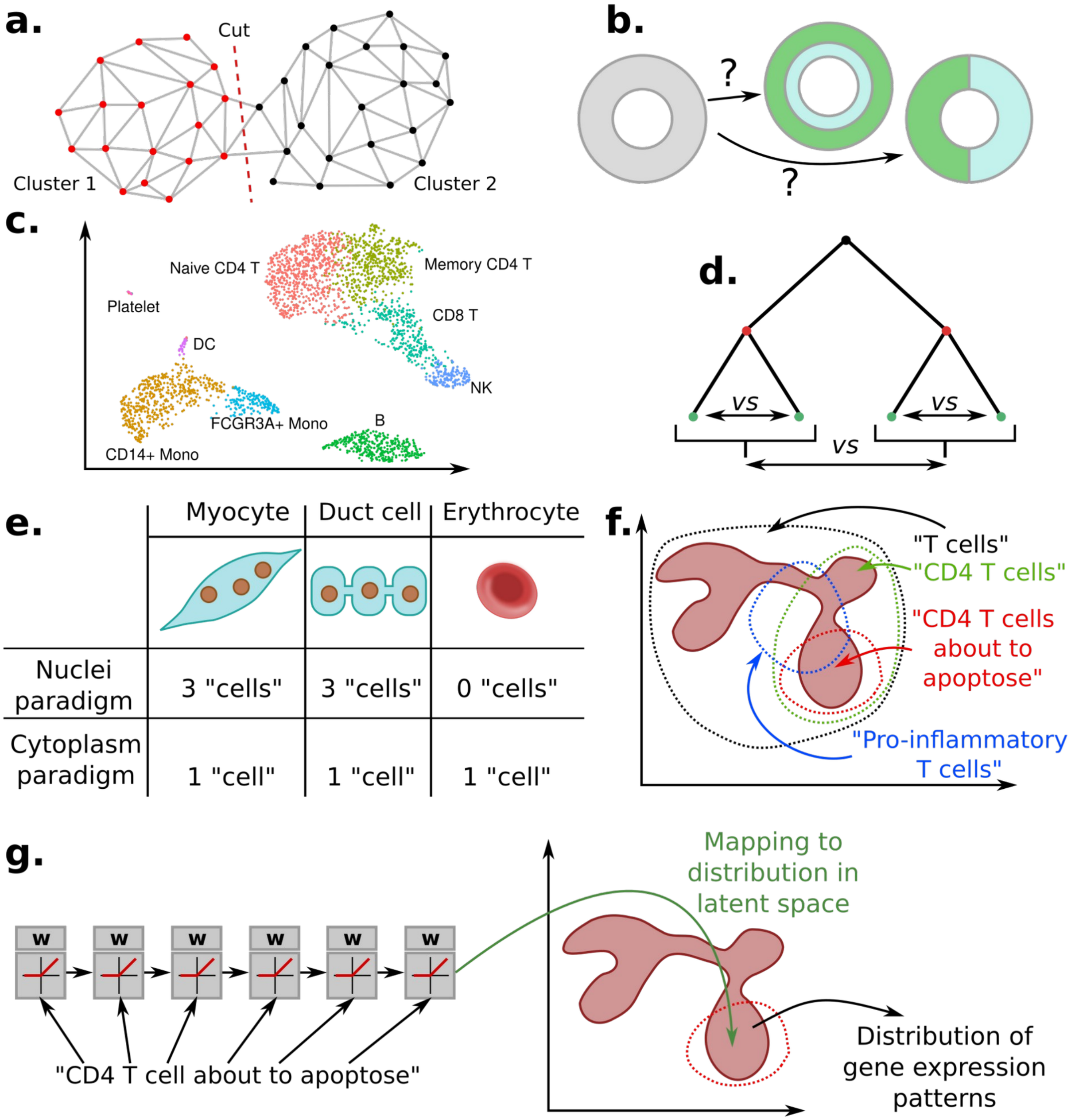

**Figure 10: Clustering, language, and cell types. (a)** Clustering can be performed by computing the best place to cut a graph. **(b)** Two ways of clustering cells in a topology; it is not clear which way is correct. **(c)** A typical clustering, with some clusters having distinct boundaries, some without. **(d)** Cell types can be viewed over a lineage tree, based on their differentiation path from progenitors. Comparisons are then mostly relevant over cell types within the same lineage. **(e)** Does a cell type refer to the nucleus, or what is inside the cell membrane? Our language is not always used consistently but it is still productive. **(f)** Natural language can hypothetically be used as a latent space, but to what extent? An NLP-based latent space need not form a tree of cell types. **(g)** Hypothetical NN architecture that can take a sequence of words, map it to a distribution in latent space, which in turn can be decoded as possible gene expression patterns.

## Data that still begs for representations

This limited review covers the most common single-cell concepts and how these can be mapped to latent representations. However, how to best represent newer types of readouts are open problems.

RNA-seq also captures genomic sequence information, for example SNPs. This can be used to separate cells into donors (for humans, not inbred mice), for example by Vireo (Huang et al. 2019). This is a simple categorical representation. Several studies, however, attempt to trace cell lineages ("lineaging") from accumulated mutations. The underlying representation is then a (lineage) tree. This representation is invalid for cells that have fused, e.g., myocytes, and rather begs for a direct acyclic graph. Because of lack of data, such representations have not yet been developed to the authors knowledge.

Spatial transcriptomics will bring the next level of challenges to the single-cell world. Combining spatial and single-cell data enables other types of statistics than presented here (Svensson et al. 2018; Liu et al. 2021). Some spatial methods measure the 3D location of individual RNA molecules, which can be more informative than just having the counts for each cell. The location is an important part of the regulation; cells may contain stress granules, which may soak up proteins and RNA to temporarily disable them. RNA may also be kept disabled for rapid activation. The size of neurons also makes the location of RNA important. Representations of data for these scenarios remain underdeveloped or non-existent; a challenge is to find suitable informative simplifications. Because 10x only recently announced a commercial technology for single-RNA molecule spatial resolution, we can expect an explosion of data analysis methods in this field.

Finally, genetic perturbations cause cells to shift within the latent spaces, or move outside what the latent space can describe if it is just based on unperturbed cells. The single-cell field is slowly moving towards being able to perturb large numbers of genes (even genome-wide) (Replogle et al. 2020, 2022; Peidli et al. 2022). How should the latent space be set up to cope with such a large amount of information? Of related interest is the prediction of the effects of perturbations (Qiu et al. 2022). If the vector field of cells in the cell state space can be measured, for example by RNA velocity(Bergen et al. 2021), or measured by metabolic labeling (Qiu et al. 2022), then this can also inform about the ideal latent space structure.

## Concluding remarks

This review has hopefully managed to portray that the matter of latent spaces, or representations, is at the heart of understanding biology at the single-cell level. Luckily, easy-to-use frameworks have been developed, which at least are good enough for testing new representations (turning it into a new analysis package further requires software engineering skills). Complex hierarchical Bayesian models (Gelman and Hill 2006) can be directly formulated using, for example, STAN (Carpenter et al. 2017). While solving them is slow in STAN, the flexibility makes up for it. Another option for solving Bayesian equations, approximately but fast, is the use of Variational Inference, which goes beyond the VAE example in this review. Bayesian equations have recently gained much traction, especially in conjunction with the use of GPUs. PyTorch (https://pytorch.org/) and Tensorflow (https://www.tensorflow.org/) are two frameworks in which it is fairly straight-forward to formulate variational inference problems. Interested readers should investigate SCVI (Gayoso et al. 2022) and scArches (Lotfollahi et al. 2022), which uses this for solving VAEs; and Cell2location (Kleshchevnikov et al. 2022) which uses this to link single-cell data to spatial transcriptomics data. Large number of free lectures on ML methods are now also available on common streaming platforms, with further examples on Github.

# Data availability

The code used to generate some of the graphs is available at Github, https://github.com/henriksson-lab/singlecell_review2023.

# Funding

I.S.M is supported by the Umeå industrial doctoral school (Företagsforskarskolan) of Umeå university, and Sartorius. J.H.is supported by Vetenskapsrådet grant number #2021-06602.

# Ethical Approval

Not applicable to this review.

# Consent to Participate

Not applicable to this review.

# Consent to Publish

Not applicable to this review.

# Conflict of interest

I.S.M is partially funded by Sartorius. Other authors declare no conflict of interest.